\documentclass[aps,pre,twocolumn]{revtex4-1}
\usepackage{graphicx}
\usepackage{amsmath,verbatim,amssymb,stmaryrd,bm}
\usepackage{hyperref}
\usepackage{xcolor}
\usepackage{comment}
\usepackage[caption = false]{subfig}

\begin{document}
	

\title{Pair-beam propagation in a magnetised plasma for modelling the polarized radiation emission from gamma-ray bursts in laboratory astrophysics experiments}
\author{Ujjwal Sinha}
\author{Naveen Kumar}\email[Corresponding author: ]{naveen.kumar@mpi-hd.mpg.de}\affiliation{Max-Planck-Institut f\"ur Kernphysik, Saupfercheckweg 1, D-69117 Heidelberg, Germany}
	
\date{\today}

\begin{abstract}
The propagation of a relativistic electron-positron beam in a magnetized electron-ion plasma is studied, focusing on the polarization of the radiation generated in this case. Special emphasis is laid on investigating the polarization of the generated radiation for a range of beam-plasma parameters, transverse and longitudinal beam sizes, and the external magnetic fields. Our results not only help in understanding the high degrees of circular polarization observed in gamma-rays bursts but they also help in distinguishing the different modes associated with the filamentation dynamics of the pair-beam in laboratory astrophysics experiments. 
\end{abstract}
\maketitle
\section{Introduction}\label{introduction}

Colliding or interpenetrating plasma flows are widely believed to be present in many astrophysical scenarios such as Supernova Remnants (SNRs), Active Galactic Nuclei (AGNs), and Pulsar Wind Nebulae (PWNs)  in the Universe. The composition of these plasma flows is dependent not only the constituents of the source but also on the energy released in these events. For example, due to the extreme energy released, \emph{e.g.} $E \sim 10^{51}$ ergs  in a small volume of radius $ \sim 10^3$ km, in  gamma-ray bursts (GRBs)~\cite{piran2005}, one can expect a copious amount of pair plasma ($e^+, e^-$) to be generated. Thus, the plasma jets emitted in these energetic events contain both hadronic and leptonic components. 

In GRBs, colliding plasma flows are generated both within the ejecta (internal shocks) and when the ejecta (external shocks) collides with the interstellar medium.  As the velocity distribution is highly anisotropic, such flows are susceptible to the Weibel or Current Filamentation instabilities (WI/CFI)~\cite{weibel1959,fried1959}. The WI/CFI mechanisms are capable of generating a strong magnetic field, of equipartition levels, transverse to the plasma flows. Thus, the WI/CFI mechanisms are believed to be responsible for the origin of such a strong magnetic field that is required to explain the power-law distribution of cosmic rays accelerated by collisionless shocks via the Fermi acceleration process~\cite{silva2003,spitkovsky2008}. In case of a magentized plasma, the magnetic obliquity with respect to the flow direction affect the shock structure and the particle acceleration~\cite{sironi2009A}. The synthetic radiation spectra obtained from the accelerated particles moving in the WI/CFI fields have been associated with the synchrotron mechanism~\cite{sironi2009, hededal2005}.  The polarization of the radiation emitted in relativistic plasma jets holds vital clues not only about the dynamical physical processes but also the composition of the relativistic jets~\cite{Anantua:2019aa}. The recent observation of polarized radiation from GRBs~\cite{mundell2013, wiersema2014, lyutikov2003}, in particular, circular polarization~\cite{wiersema2014}, has invoked immense interest to revisit the theoretical models describing such scenarios~\cite{medvedev1999, gruzinov1999}. Polarized radiation indicates the presence of large scale ordered magnetic fields~\cite{mundell2013}. Recently, we have shown that the asymmetric dissipation of kinetic energy between the species of an electron-positron beam propagating in a magnetized electron-ion plasma can produce circularly polarized radiation~\cite{sinha2019}. Understanding the mechanism of the generation of polarized radiation helps in understanding the magnetic field configuration, plasma composition and the geometry of the emitting region of GRBs~\cite{sinha2019,Sinha2018ab,*Sinha:2015aa}.

With the availability of intense lasers, it has been possible to generate colliding flows mimicking the astrophysical scenarios in a laboratory set-up. These experiments have reported the generation of current filaments and equipartition magnetic field~\cite{fox2013, huntington2015}. The generation of a relativistic high-density charge-neutral electron-positron plasma in the laboratory has opened up the possibility to investigate the physics of GRBs in laboratory astrophysics experiments~\cite{Sarri:2015ab}. In these experiments, the proton radiography technique is primarily employed to investigate the structure of the current filaments and the magnetic field strength~\cite{fox2013, huntington2015, fiksel2014}.  Although this technique has deeply increased our understanding of the physics of interpenetrating plasma flows, the radiation and polarization signatures can further help in understanding the dissipation mechanism at the kinetic scale~\cite{sinha2019,Sinha2018ab}.

As mentioned before, we have recently studied the polarized radiation generation  from a  charge-neutral electron-positron ($e_b^-,e_b^+$) beam propagating in a magnetized ambient electron-proton ($e^-,p^+$) plasma  by means of multi-dimensional PIC simulations and a newly developed code that post-processes the particle trajectories from PIC simulations~\cite{sinha2019}. In this paper, we extend our previous results to ultra-relativistic regime and also study the influence of transverse and longitudinal beam sizes on the polarization of the generated radiation. By varying the Lorentz factor we {cover} a broad range of magnetization parameter $\sigma_0$ which is a scale-invariant parameter that facilitates comparison between laboratory astrophysics experiments and astrophysical observations using the scaling and self-similarity arguments. Moreover, the influence of the longitudinal and transverse beam-sizes on the polarization of the radiation also provides a new way to detect these plasma instabilities in laboratory astrophysics experiments. The remainder of this paper is organized as follows: In sections \ref{pic_params} and \ref{casper} we respectively describe our PIC simulation parameters and the numerical technique employed to model the radiation and polarization from particle trajectories. In Sec.~\ref{gamma}, we show the effect of the initial beam to plasma density ratios and the magnetization of the ambient plasma on the polarization of the radiation. The radiation is shown to be emitted via the synchrotron mechanism. The combined effect of transverse and longitudinal modes of perturbation on the polarization of radiation is studied in Sec.~\ref{long_beam} by varying the longitudinal length of the beam. In Sec.~\ref{grbs}, we describe the relevance of our studies for the recently observed circular polarization in GRBs.

\section{PIC simulation details and choices of beam-plasma parameters}\label{pic_params}
We carry out three-dimensional (3D) PIC simulations using the open-source code SMILEI~\cite{derouillat2018}. We employ a relativistic charge neutral electron-positron ($e_b^-,e_b^+$) beam moving along the $x$ direction through a uniform ambient magnetized electron-proton ($e^-,p^+$) plasma. The ($e_b^-,e_b^+$) beam has a Gaussian density profile in all directions. A moving (in $x$ direction) simulation window with resolutions of [$16\times 8\times 8$] cells per $c/\omega_p$ and absorbing boundary conditions for particles and fields in the transverse direction were used. Here, $\omega_p=(4\pi n_0e^2/m_e)^{1/2}$ is the electron plasma frequency of the ambient ($e^-,p^+$) plasma, $n_0$ the electron density, $e$ and $m_e$ the electronic charge and mass respectively and $c$ is the velocity of light in vacuum. The proton mass $m_p = 1836m_e$. A timestep of $\Delta t=0.0295\,\omega_p^{-1}$ and $4$ particles per cell per species were used. We also performed limited set of simulations with larger number of particles per cell (32 and 100), obtaining essentially the same results. The ambient plasma has a temperature of $\sim 5$ keV and was magnetized with a uniform magnetic field $\textbf{B}_0=B_0\hat{\textbf{e}}_x$, where $B_0$ is normalized to $m_e\omega_pc/e$. The magnetization parameter is defined as {$\sigma_0=B_0^2/(4\pi\gamma_0 m_en_{b0}c^2)$}~\cite{sironi2009A}, where {$\gamma_0$ is the initial relativistic factor and $n_{b0}$ is the peak density} of the ($e_b^-,e_b^+$) beam. In order to quantify the role of the plasma instabilities on the beam propagation and consequently radiation generation, we focus on first investigating the effect of large Lorentz factor {$\gamma_0 \ge 100$} on the polarized radiation generation. Then we focus on investigating the influence of transverse and longitudinal sizes of the pair-beam on the polarized radiation generation.  This not only helps in deciphering the role of the beam sizes on polarized radiation generation, but it also provides an useful scaling that is essential for experimentally investigating the dynamics of the WI/CFI of a pair-beam in laboratory astrophysics experiments. Our choices of parameter not only covers the broad parameter regimes for upcoming FACET-II~\cite{Yakimenko:2019aa} facility but is also  relevant for scaling our results to astrophysical observations of GRBs.

\section{Computation of the radiation and its polarization by extracting particle trajectories from PIC simulations}\label{radiation}

PIC simulations allow us to understand the motion of the charged particles in self-generated plasma fields and are extensively used as a powerful tool to study kinetic effects in high energy density physics (HEDP) scenarios arising in intense laser-plasma interactions. Strong electromagnetic fields are produced in these scenarios due to generation of density and temperature gradients (Biermann mechanism)~\cite{biermann1950} or instabilities (Weibel, current filamentation or two-stream instability)~\cite{weibel1959, fried1959}. As the charged particles move in these fields at relativistic velocities, their trajectories bend and they experience acceleration perpendicular to their velocity. The resulting emission is directed along the particle velocity and is confined within a cone of half angle $1/\gamma$, where $\gamma$ is the Lorentz factor of the particle~\cite{jackson1999}. The radiated fields have frequency bandwidths spanning several orders of magnitude with the cut-off frequency of radiation scaling as  $\omega_c \propto \gamma^3$. For the radiation generation due to CFI/WI mechanism, the important spatial scale is the plasma skin-depth. The ability of PIC codes to capture such radiation is limited by the grid resolution. Resolving the radiated fields directly on the PIC grid requires extremely large computational resources. However, as the PIC codes keep track of the individual particle trajectories at all time-steps and during the whole simulation duration, a practical and efficient method is to post-process the position and momenta of the particles over time to calculate the radiated fields at a fixed point of observation~\cite{hededal2005, martins2009}. If such information is available for a selection of particles, then the entire radiation spectrum can be obtained.

When a relativistic ($e_b^-,e_b^+$) beam propagates through a plasma, a transverse perturbation in magnetic field allows the WI/CFI mechanisms to grow and generates current filaments. The beam filamentation represents the macroscopic time-varying charge and current densities of the beam species. The charged particles are confined in these filaments and have associated magnetic field surrounding them with azimuthal symmetry. This magnetic field spatially confines the particles. The motion of the charged particles in these filaments is such that their maximum displacement in the transverse plane is of the order of the filament transverse size. The azimuthal field bends the particle trajectories when they reach the filament edge. This produces a strong curvilinear motion with the acceleration perpendicular to the velocity. Such a motion allows the particles to radiate. The expression of the radiated  field,  $\textbf{E}_{\textrm{rad}}(t^\prime)$, emitted by a particle with charge, -$e$, is given as~\cite{jackson1999},
\begin{align}\label{fld_vec}
\textbf{E}_{\textrm{rad}}(t^\prime) &= \Bigg[ \frac{e}{c}\frac{\textbf{n}\times\{(\textbf{n}-\boldsymbol{\beta})\times \dot{\boldsymbol{\beta}}\}}{(1 - \boldsymbol{\beta}\cdot\textbf{n})^3R} \Bigg]_{ret},\nonumber \\
&=\Bigg[\frac{e}{c}\frac{(\boldsymbol{\beta}\cdot \textbf{n} - 1)\dot{\boldsymbol{\beta}} + \textbf{n}\cdot\dot{\boldsymbol{\beta}}(\textbf{n} - \boldsymbol{\beta})}{(1 - \boldsymbol{\beta}\cdot\textbf{n})^3R}\Bigg]_{\textrm{ret}},
\end{align}
where $R(t) = x - \bm{n} \cdot \bm{r}(t)/c$ is the distance from the charge particle to the observation point in the detector, $x$ is the distance from the origin to the observation point, $r(t)$ is the moving charge trajectory, and $\bm{\beta} = d\bm{r}/dt$ and $\bm{\dot{\beta}}=d\bm{\beta}/d t$, are the velocity and acceleration normalised to $c$, respectively. In addition, the subscript \emph{ret} means that quantities are evaluated at the retarded time $t^{'} = t + R(t)/c$. 
Eq.\eqref{fld_vec} establishes the relation between the direction of the radiated electric field and the particle velocity and acceleration vectors. The radiated field is either linearly or circularly polarized depending upon whether the velocity and acceleration vectors of the charged particle oscillate either in one direction or rotate in the transverse plane as it propagates through the plasma. The radiation is unpolarized if the velocity and acceleration vectors randomly change directions. Eq.\eqref{fld_vec} thus qualitatively explains the generation of polarized radiation  in laboratory and astrophysical scenarios.

\subsection{Post-processing code CASPER}\label{casper}
To compute the radiation, we have developed a new radiation post-processing code CASPER, which stands for Coherence and Spectral Postprocessor for Emitted Radiation~\cite{sinha2019}. CASPER employs a two-dimensional virtual detector kept at a distance far from the simulation box. CAPSER can be directly used to post-process the radiation from open-source PIC codes such as SMILEI and EPOCH. For the purpose of the computation, we employ the Fourier transformed radiated electric field away from simulation box on the grid-points  of the virtual detector. The expression for the radiated field reads,
\begin{multline}\label{rad_E}
\textbf{E}_{\mathrm{rad}}(\omega)=\left(\frac{e^2}{2\pi c^2}\right)^{1/2}\int_{-\infty}^{\infty}
\frac{\textbf{n}\times [(\textbf{n}-\boldsymbol{\beta})\times \dot{\boldsymbol{\beta}}]}{R(t^{'})(1-\boldsymbol{\beta}\cdot \textbf{n})^2} \\e^{i\omega(t^\prime-\textbf{n}\cdot \textbf{r}(t^\prime)/c)} dt^\prime,
\end{multline}
where $\textbf{r}(t^\prime)$ is the instantaneous position of the particle. For a specified particle motion $\textbf{r}(t^\prime)$, the momentum $\textbf{p}(t^\prime)$ is obtained from PIC simulations. Thus, one  can compute $\boldsymbol{\beta}$ and $\dot{\boldsymbol{\beta}}$ at every timestep and the integral is evaluated for a range of emitted frequency $\omega$. This allows us to obtain the spectra of the radiated fields beyond the resolution limit imposed by the PIC simulations. To obtain the spectrum from a population of charged particles, the amplitude $\textbf{E}_\mathrm{rad}(\omega)$ obtained for each one of them is coherently added. Having the information about the radiated electric field enables us to compute the Stokes parameters and determine the polarization state of the electromagnetic radiation. In terms of the unit vectors $\boldsymbol{\epsilon}_1$ and $\boldsymbol{\epsilon}_2$ perpendicular to the direction of the radiation observation, the Stokes parameters are given as~\cite{jackson1999},
\begin{subequations}\label{stokes}
	\begin{align}
	\label{eq: 4a}
	s_0 &= |\boldsymbol{\epsilon}_1\cdot\textbf{E}_\mathrm{rad}|^2 + |\boldsymbol{\epsilon}_2\cdot\textbf{E}_\mathrm{rad}|^2, \\
	\label{eq: 4b}
	s_1 &= |\boldsymbol{\epsilon}_1\cdot\textbf{E}_\mathrm{rad}|^2 - |\boldsymbol{\epsilon}_2\cdot\textbf{E}_\mathrm{rad}|^2, \\
	\label{eq: 4c}
	s_2 &= 2\hspace{1mm}\mathrm{Re}[(\boldsymbol{\epsilon}_1\cdot\textbf{E}_\mathrm{rad})^*(\boldsymbol{\epsilon}_2\cdot\textbf{E}_\mathrm{rad})],\\
	\label{eq: 4d}
	s_3&= 2\hspace{1mm}\mathrm{Im}[(\boldsymbol{\epsilon}_1\cdot\textbf{E}_\mathrm{rad})^*(\boldsymbol{\epsilon}_2\cdot\textbf{E}_\mathrm{rad})].
	\end{align}
\end{subequations}
Eqs.\eqref{stokes} allow us to calculate the total radiated intensity as a linear combination of the intensities measured in the orthogonal polarization directions. The parameter $s_0$ describes the total intensity of the radiation, $s_1$ describes the excess of linear polarization in the horizontal direction over the vertical direction, $s_2$ describes the excess of linear polarization at an angle of $45^o$ with the horizontal over the one at an angle of $135^o$, and $s_3$ describes the excess of right circularly polarized light over left circularly polarized light. As the linear polarization is described by two Stokes parameters $s_1$ and $s_2$, the total linearly polarized flux is given by,
\begin{equation}
P_\mathrm{lflux} = \sqrt{s_1^2 + s_2^2}.
\end{equation}
The degrees of linear and circular polarizations are given by $P_\mathrm{lin}=P_\mathrm{lflux}/s_0$, and $P_\mathrm{circ}=s_3/s_0$, respectively. To derive the total linear ($\langle P_\mathrm{lin}\rangle$) or circular polarization ($\langle P_\mathrm{circ}\rangle$) components of the entire radiation, we first weigh the degree of the corresponding polarization with the intensity and integrate it over the entire area of the detector and the frequency bandwidth. If $A$ is the total detector area, then
\begin{subequations}\label{pol_degree}
	\begin{align}
	\label{erad}
	\langle \mathcal{E}_\mathrm{rad}\rangle &= \frac{\iint s_0 dA\, d\omega}{\iint dA\,d\omega},\\
	\label{plin}
	\langle P_\mathrm{lin}\rangle &= \frac{\iint P_{\textrm{lflux}}dA\,d\omega}{\iint s_0 dA\, d\omega},\\
  	\label{pcirc}
    \langle P_\mathrm{circ}\rangle &= \frac{\iint s_3 dA\, d\omega}{\iint s_0 dA\,d\omega}.
    \end{align}
\end{subequations}
We can integrate Eqs.\eqref{pol_degree} over the detector area, A, alone to obtain the spectra of the radiation ($\langle \mathcal{E}_\mathrm{rad}\rangle_\mathrm{A}$) and the flux corresponding to linear ($\langle {P}_\mathrm{lin}\rangle_\mathrm{A}$) and circular ($\langle {P}_\mathrm{circ}\rangle_\mathrm{A}$) polarization. The obtained spectra then allows to understand frequency bandwidths contribution to the  specific polarization state. Similarly, Eqs.\eqref{pol_degree} can be integrated over $\omega$ to obtain the spatial distribution of the radiation and polarization on the detector area.

\subsection{Scaling for the degree of circular polarization}\label{scaling}

For a distribution of particles, the total radiation emitted is a coherent superposition of the contributions coming from all the particles. In case of a magnetized ambient plasma, the beam particles move in a combination of the initial magnetic and filamentation induced fields. Since the magnetic field associated with the WI/CFI mechanism is azimuthally symmetric, the particles bounce in all possible directions in the transverse plane resulting in an isotropically distributed velocity and acceleration vectors. The initial axial field, however, {tries} to rotate the particle in the transverse plane inducing a finite degree of coherence in the velocity and acceleration vectors. As mentioned before, a coherent motion is essential for polarized radiation. Thus, the polarization of the radiation depends on the competition between the effect of initial magnetization and the WI/CFI fields on the particle trajectories. The WI/CFI fields deflect the particles at the spatial scales of the filament transverse size, $\lambda_\mathrm{CFI}$. Whereas the effect of the initial magnetization is associated with the Larmor radius, $r_\mathrm{L}=\gamma m\beta/qB$. Thus, the ratio $\eta = r_\mathrm{L}/\lambda_\mathrm{CFI}$ determines the competition between the initial magnetic field and the {WI/CFI} generated magnetic field. When $\eta >1$ \emph{i.e} $r_\mathrm{L}>\lambda_\mathrm{CFI}$, the transverse velocity and acceleration vectors are influenced by the small-scale {WI/CFI} generated fields which causes a degree of de-polarization on the radiation generation. Consequently, the radiation contains a smaller component of polarized light. Similarly, when $r_\mathrm{L}<\lambda_\mathrm{CFI}$, the fields generated due to the WI/CFI mechanisms are dominant on larger spatial scales and hence the initial magnetization can cause the generation of polarized light. For a relativistic particle the $r_\mathrm{L}$ increase with $\gamma$. Hence, for high $\gamma$ the ratio $\eta$ increases, resulting in a smaller component of  polarized radiation generation. The external magnetic field is ordered while the   magnetic field  due to the {WI/CFI} mechanisms is inhomogeneous. Thus, it is essential to study the effect of large $\gamma$ (as in astrophysical flows) and the transverse and longitudinal sizes of the pair-beam.  Keeping these in mind, one can formulate a scaling, based on the synchrotron radiation formalism, that estimate the maximum amount of polarized radiation generation due to the external magnetic field from PIC simulations. 

\begin{figure}
	\includegraphics[width=0.5\textwidth,height=0.40\textwidth]{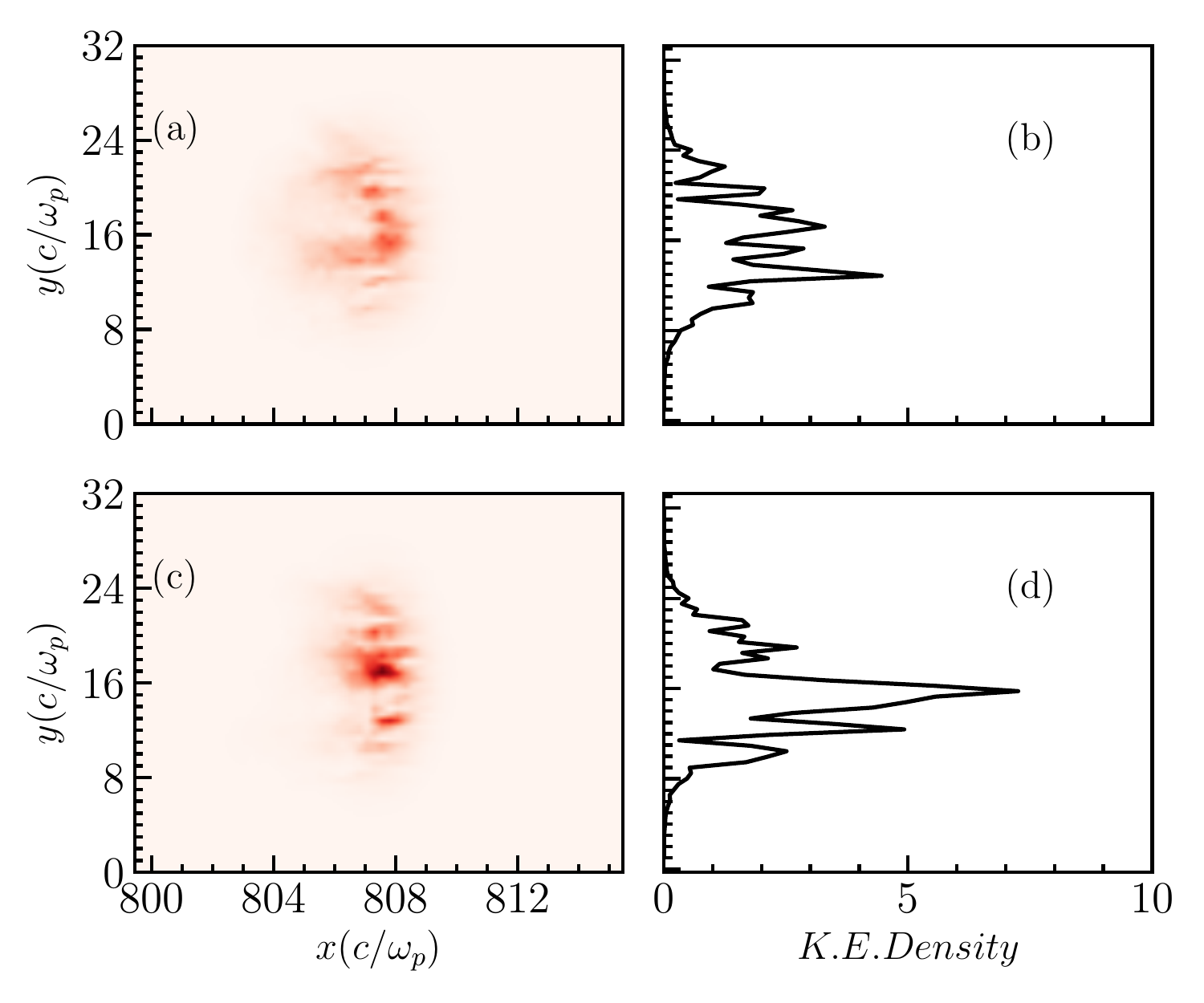}
	\caption{Spatial distribution of the kinetic energy density \textcolor{black}{(normalized with $e^2/m_e^2\omega_p^2 c^2$)} of the beam electron (a) and positron (c) in the $xy$ plane at $z = 16 c/\omega_p$ and $t = 800\,\omega_p^{-1}$. The initial beam $\gamma_0 = 50$ and $n_\mathrm{b0}/n_0 = 1.0$. The beam length $L_x = 6\,c/\omega_p$ and $L_y = L_z = 25.6\, c/\omega_p$. Panels (b) and (d) show the lineouts at $x = 808\, c/\omega_p$ for corresponding distributions in panels (a) and (c), respectively.}
	\label{ene_den_map}
\end{figure}

To compute this, we proceed with the scenario of a charged particle moving in an instantaneous circular motion. The radiation is beamed in a narrow cone in the direction of the velocity vector and is seen by the observer as a short pulse of radiation. For a relativistic particle, the angle $\theta$ (between the $x$-axis and the unit vector $\mathbf{n}$) at which the emission takes place is small. Using this approximation, the transverse components {$E_1\boldsymbol{\epsilon}_1$}  and $E_2\boldsymbol{\epsilon}_2$ of Eq.\eqref{rad_E} takes the form,

\begin{subequations}\label{rad_comp}
	\begin{align}
		E_1(\omega) &=\Bigg(\frac{2 e^2}{\pi c^2 R^2}\Bigg)^{1/2}\frac{\omega \rho}{c}\Bigg(\frac{1}{\gamma^2}+\theta^2\Bigg)\frac{1}{\sqrt{3}}K_{2/3}(\xi),\\
		E_2(\omega) &=\pm i\Bigg(\frac{2e^2}{\pi c^2 R^2}\Bigg)^{1/2}\frac{\omega \rho}{c}\theta\Bigg(\frac{1}{\gamma^2}+\theta^2\Bigg)^{1/2}\frac{1}{\sqrt{3}}K_{1/3}(\xi),
	\end{align}
\end{subequations}
where $\xi = (\omega\rho/3c)(1/\gamma^2 + \theta^2)^{3/2}$ and $\rho$ is the instantaneous radius of curvature, $\theta$ is the half-angle of the cone emission of radiation, $+$ is for $e_b^+$ and $-$ is for $e_b^-$. The radiation is {partially} circularly polarized for $\theta\neq 0$. Using Eqs.\eqref{stokes} and \eqref{rad_comp}, one can calculate the Stokes parameters $s_0$ and $s_3$. Integrating $s_3$ and $s_0$ over the entire frequency domain and the solid angle we obtain the circularly polarized photon flux $P_\mathrm{cflux}=7.2(e/c)^2\omega_0\gamma^4$ and the radiated energy $\mathcal{E}_{rad}=8.556(e/c)^2\omega_0\gamma^4$ from a single particle{, where $\omega_0$ is the instantaneous Larmor frequency}.

\textcolor{black}{As mentioned before, during the beam transportation in an ambient plasma, an initial transverse magnetic field perturbation exerts the Lorentz force which is in opposite directions for both species due to the two species having opposite charges. Since the pair beam has a transverse size larger than the plasma skin-depth $d_s = c/\omega_p$, the return plasma current can penetrate and flows inside the pair beam. This results into smaller beam filaments formation. These individual filaments generate their own magnetic fields which superimpose with the initial magnetic field perturbation. This leads to the growth of the magnetic field perturbation and consequently filamentation of the beam. In this process, a fraction of beam energy is converted into the strong quasi-static transverse magnetic field generation. Moreover, the beam electrons ($e^-_b$) repel and the beam positrons ($e^+_b$) attract the background plasma electrons, respectively. Combination of these processes causes an asymmetry in the Lorentz factor of both beam species, leading to the difference in their energy spectrum, as seen in Fig.\ref{ene_den_map}.} \textcolor{black}{For the purpose of calculating the radiation and degrees of polarizations, one can employ the single particle synchrotron emission, albeit generalizing it for a distribution of beam particles}. For a pair ($e_b^+,e_b^-$) beam with distribution $f(\gamma)_\alpha$, where $\alpha=[e_b^+, e_b^-]$, the resultant degree of circular polarization, \textcolor{black}{$\langle P_\mathrm{circ}\rangle = (\langle P_\mathrm{cflux}\rangle_{e_b^-}-\langle P_\mathrm{cflux}\rangle_{e_b^+})/(\langle \mathcal{E}_\mathrm{rad}\rangle_{e_b^-}+\langle \mathcal{E}_\mathrm{rad}\rangle_{e_b^+})$}, where $\langle P_\mathrm{cflux}\rangle = \int P_\mathrm{cflux}f(\gamma)_\alpha d\gamma/\int f(\gamma)_\alpha d\gamma$, yields the scaling for $\langle P_\mathrm{circ}\rangle$ as,
\begin{equation}\label{scaling}
	\langle P_\mathrm{circ}\rangle = 0.8415\frac{\langle\gamma^4\rangle_{e_b^-}-\langle\gamma^4\rangle_{e_b^+}}{\langle\gamma^4\rangle_{e_b^-}+\langle\gamma^4\rangle_{e_b^+}}\,.
\end{equation}
\noindent
For a charge neutral ($e_b^+,e_b^-$) beam, it is important to note that the $e_b^+$ and $e_b^-$ produce Larmor gyrations with opposite helicities.  When {their} energies are similar, there is no circularly polarized radiation generation. However, it has been shown, when the ($e_b^+,e_b^-$) beam propagates through an ($e^-,p^+$) ambient plasma, the $e_b^+$ accelerate and the $e_b^-$ decelerate leading to a difference in their final kinetic energies~\cite{sinha2019}. The degree of circular polarization in this case is given by Eq.\eqref{scaling}. Thus, the $\langle P_\mathrm{circ}\rangle$ depends on both $\eta$ and the dissipation in species energies. \textcolor{black}{The degree given by Eq.\eqref{scaling} is the maximum value and one can expect to observe lower values of the circular polarization due to the de-polarizing effects of the {WI/CFI} generated magnetic field}. The scaling with respect to beam Lorentz factor has been shown to agree with the PIC simulation results, which also confirm the de-polarization effects arising due to WI/CFI generated magnetic field~\cite{sinha2019}.

\section{PIC simulation results}\label{picresults}

 As mentioned before, we begin with first showing the results on pair beam with varying Lorentz factors. Afterwards we examine the role of pair-beam spot-sizes and longitudinal length on the polarized radiation {generation} while keeping the Lorentz factor fixed. {For cases where $\gamma_0 \ge 100$, we performed simulations both with and without the radiation reaction force accounted in the plasma dynamics}. \textcolor{black}{Nevertheless, on turning off the radiation reaction force we did not notice any difference in the CFI/WI field strengths and the kinetic energy distribution of the beam species. This is expected as the plasma density used in our simulation is rather low, $n_0\sim 10^{16}$cm$^{-3}$. Consequently, the CFI/WI mechanisms do not generate a strong quasi-static magnetic field, and as a result, the radiation losses are within the Thomson scattering regime.}

\subsection{Effect of beam Lorentz factor $\gamma$ on polarized radiation generation}\label{gamma}

\begin{figure}
	\includegraphics[width=0.425\textwidth,height=0.24\textwidth]{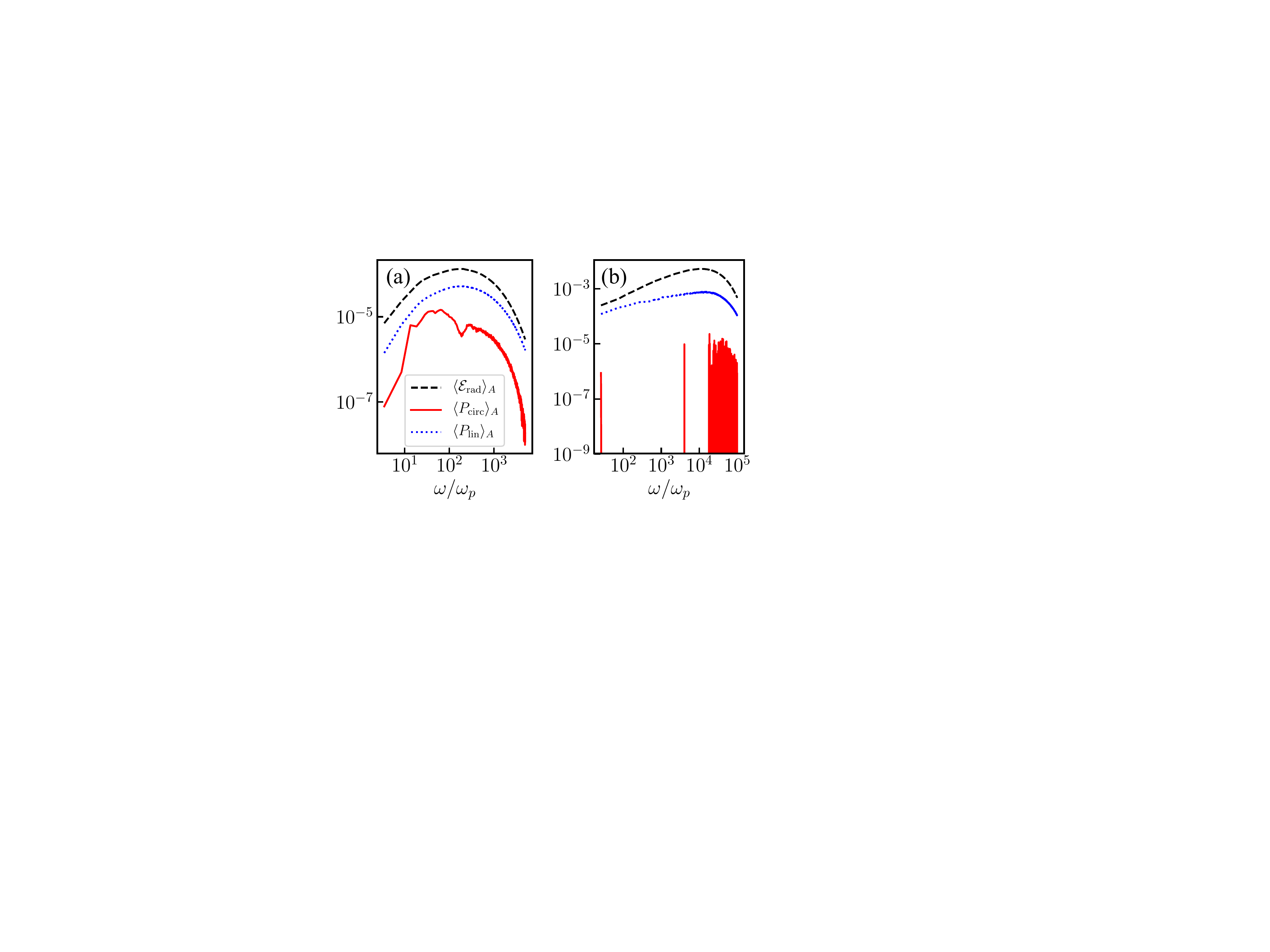}
	\caption{(a) Spatially averaged spectra of total radiated energy, linearly and circularly polarized radiation fluxes for a pair ($e_b^+,e_b^-$) beam with {$\gamma_0=100$} and $n_{b0}/n_0 = 0.1$. (b)  for  {$\gamma_0=1000$} and $n_{b0}/n_0 = 0.1$. \textcolor{black}{The radiated energy $\mathcal{E}_{\textrm{rad}}$ is normalized by $4\pi^2 c/e^2$}.}
	\label{fig3}
\end{figure}

\begin{figure}
	\includegraphics[width=0.5\textwidth, height=0.3\textwidth]{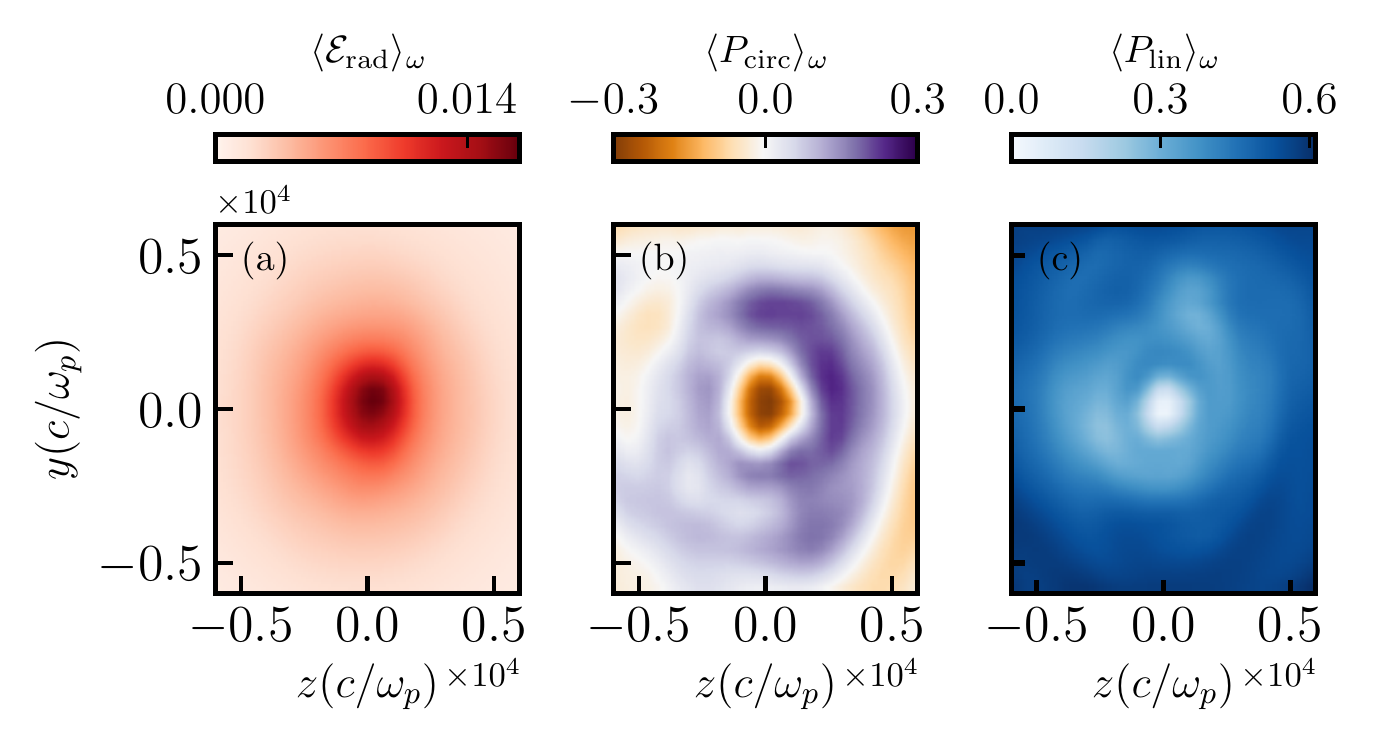}
	\caption{(a) The frequency averaged radiated energy, (b) degree of circular polarization, and (c) degree of linear polarization  for a pair ($e_b^+,e_b^-$) beam with {$\gamma_0=100$} and $n_{b0}/n_0 = 0.1$.}
	\label{fig4}
\end{figure}
\begin{figure}
	\includegraphics[width=0.5\textwidth, height=0.3\textwidth]{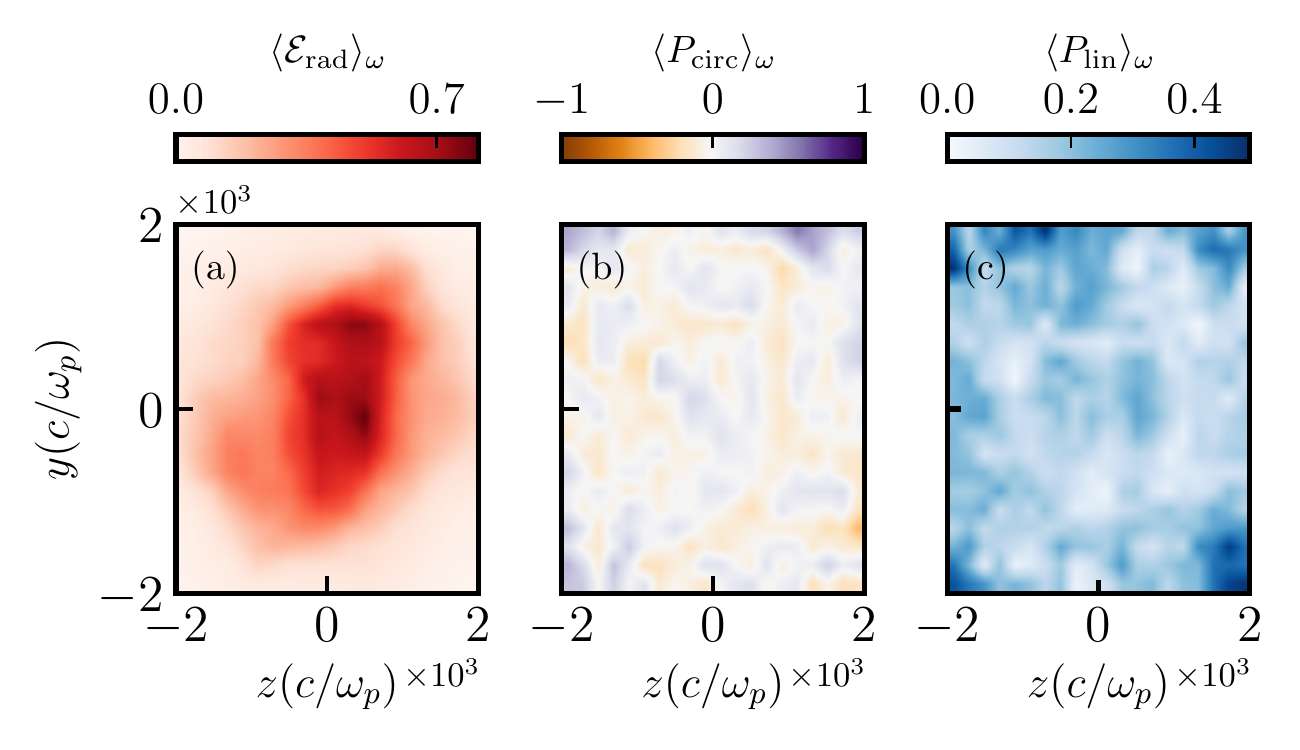}
	\caption{(a) The frequency averaged radiated energy, (b) degree of circular polarization, and (c) degree of linear polarization for a pair ($e_b^+,e_b^-$) beam with {$\gamma_0=1000$} and $n_{b0}/n_0 = 0.1$.}
	\label{fig5}
\end{figure}
\begin{figure}
	\includegraphics[width=0.42\textwidth,height=0.22\textwidth]{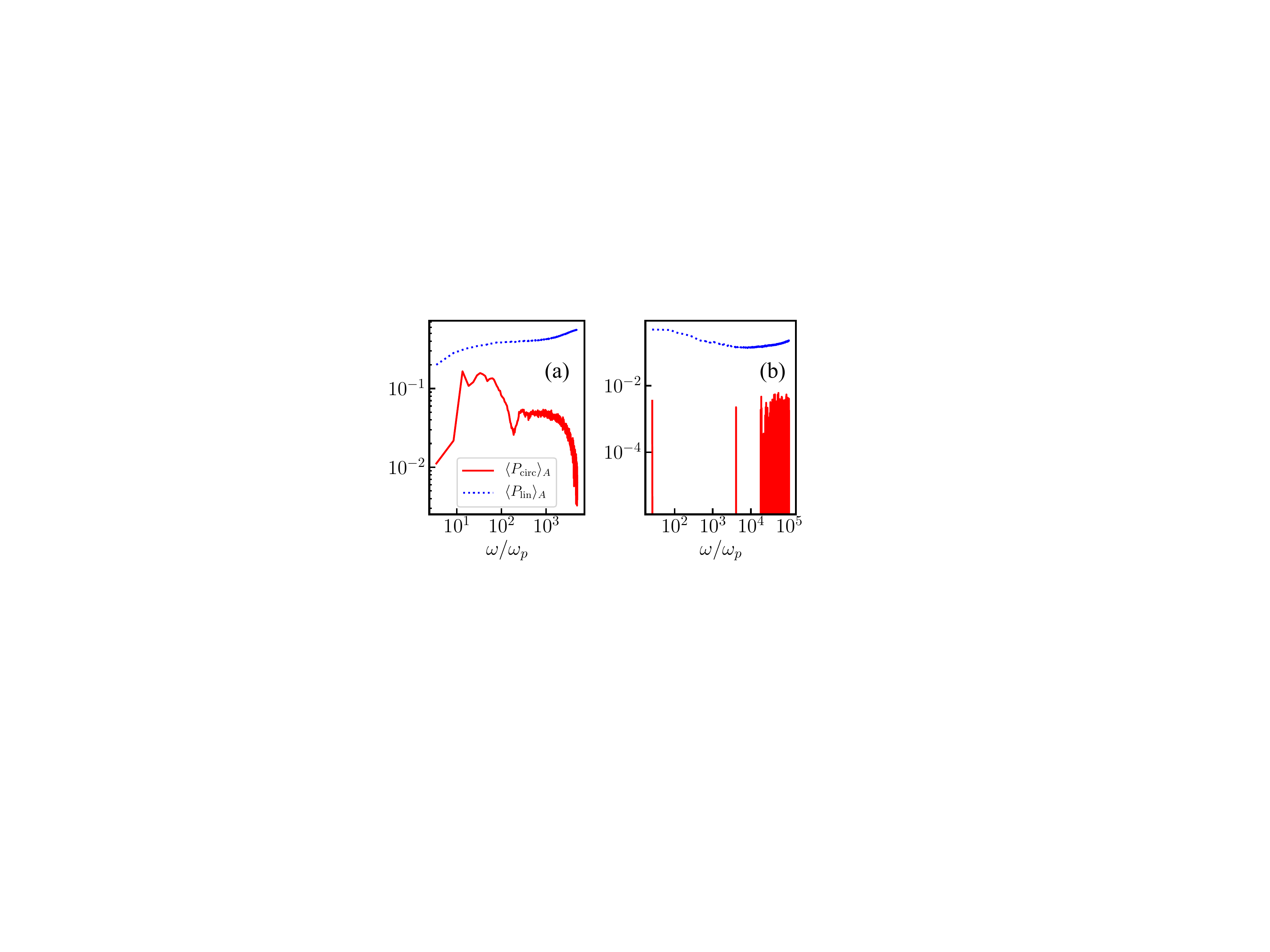}
	\caption{(a) Spatially averaged degrees of linear and circular polarizations for a pair ($e_b^+,e_b^-$) beam with {$\gamma_0=100$} and $n_{b0}/n_0 = 0.1$. (b) {Spatially averaged degrees of linear and circular polarizations for a pair ($e_b^+,e_b^-$) beam with {$\gamma_0=1000$} and $n_{b0}/n_0 = 0.1$.}}
	\label{fig6}
\end{figure}

\begin{figure}
	\includegraphics[width=0.5\textwidth,height=0.32\textwidth]{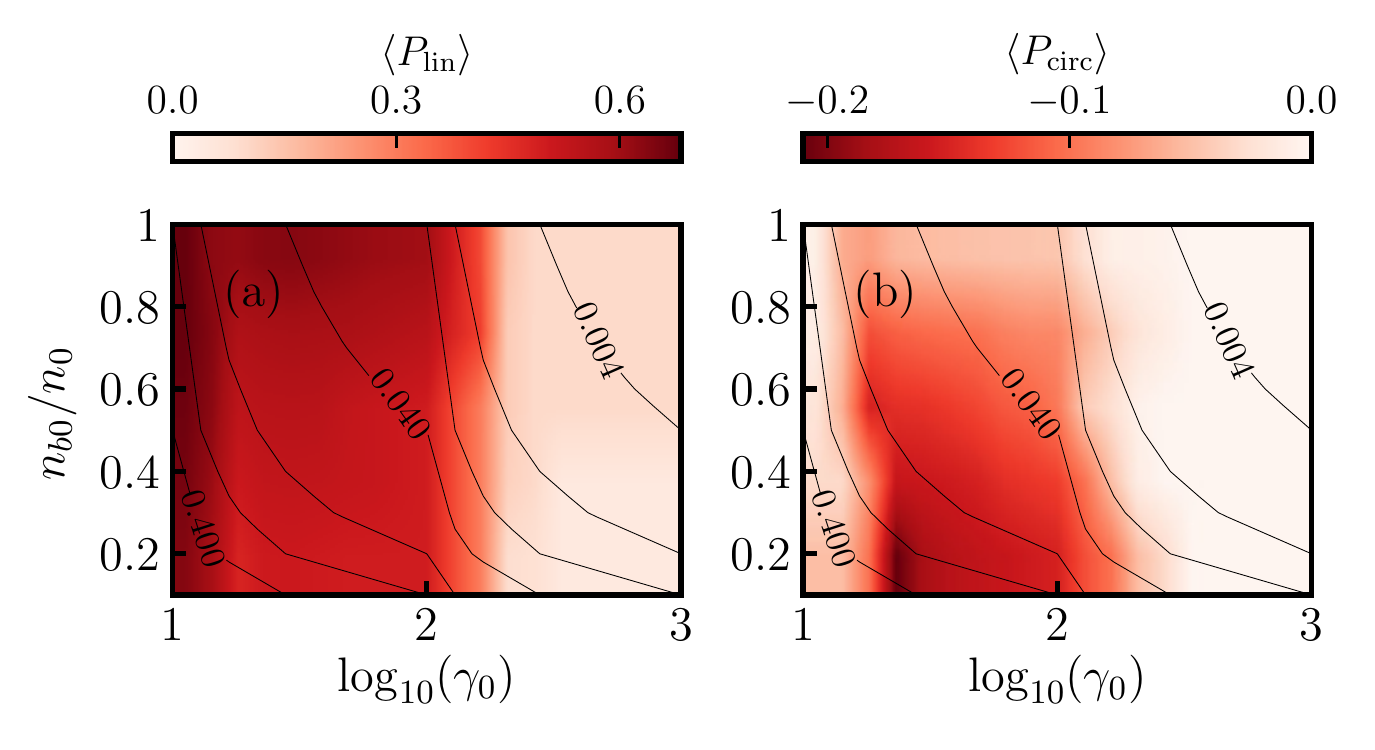}
	\caption{\textcolor{black}{Parameter maps (a) for the degree of linear polarization, and (b) and the degree of circular polarization for $n_{b0}/n_0=[0.1-1.0]$. The magnetization parameter $\sigma_0$ is depicted by the overlaid contour lines on the parameter maps and is in the range $\sigma_0=[0.002 - 2.0]$.}}
	\label{fig7}
\end{figure}

We performed simulations for a pair ($e_b^-,e_b^+$) beam of size {$L_y = L_z = 25.6\, c/\omega_p$ and $L_x = 6\, c/\omega_p$} {with a Gaussian density profile in each direction and a FWHM of one-third of the corresponding length in respective directions}. We vary Lorentz factors {$\gamma_0$} in the range, $\gamma_0=[10-1000]$. {The simulation window size is \textcolor{black}{$\Delta x\times \Delta y\times \Delta z=[16\times 32\times 32](c/\omega_p)^3$.} }We also change the density ratio, $n_{b0}/n_0=[0.1-1]$. A constant initial magnetic field $B_0 = 2$ is applied which corresponds to $\sigma_0$ in the range {$\sigma_0=[0.002 - 2.0]$}. This range of the magnetization parameter can be relevant for scaling our results to astrophysical observations of GRBs. \textcolor{black}{For the purpose of computing the radiation, we extract the trajectories of $1000$ beam $e_b^-$ and $e_b^+$ directly from the PIC simulations.} The virtual detector lies in the $yz$ plane and is kept at a distance of {$10^5 c/\omega_p$} along $x$ to ensure that the radiation is computed in the far field. {As the angular spread of the radiation is inversely proportional to the $\gamma$, the detector size was varied for different $\gamma$ such that the radiated fields are spatially well resolved.} The detector size along $y$ and $z$ is resolved using $50$ grid points in each  directions. {The radiation frequency bandwidth varies with $\gamma$}, hence the frequency range for each of the simulation case is chosen such that the entire radiation spectrum is captured. Figure \ref{fig3} shows the spatially averaged spectrum of the radiated energy and linearly and circularly polarized photon fluxes for beams with initial {$\gamma_0=[100,1000]$} at a density ratio $n_{b0}/n_0=0.1$. {The frequency ranges for these cases are $\omega=[1-5000]\omega_p$ and \textcolor{black}{$\omega=[1-10^5]\omega_p$}, respectively. The frequency axis is resolved with a step-size of $\Delta\omega = 2\omega_p$ for both cases.} For {$\gamma_0=100$}, the radiation spectrum shows significant amount of both linear and circular polarizations. While for {$\gamma_0=1000$}, the polarized component of the radiation decreases. Though the fraction of  linear polarization is still significant, the amount of circular polarization is vanishingly small in this case ($\gamma_0=1000$). Figs. \ref{fig4} and \ref{fig5} show the corresponding spatial distribution of the frequency averaged radiated energy, and the degrees of linear and circular polarizations for {$\gamma_0=100$} and {$\gamma_0=1000$}, respectively. The transverse size of the beam filaments due to the {WI/CFI} is similar in both cases, however the Larmor radius $r_L$  vary strongly due to $\gamma$ factor. Thus, for {$\gamma_0=1000$}, we get $\eta > 1$. This suggests the local dominance of the {WI/CFI} generated magnetic field over the external magnetic field for the polarized radiation generation. Consequently, the  fraction of polarized light decreases for higher Lorentz factor ({$\gamma_0=1000$}).  A small degree of linear polarization can be attributed to the small fraction of the particles belonging to the  the beam distribution satisfying $\eta < 1$. Fig.\ref{fig6} shows {the spatially averaged degrees of linear and circular polarizations}, and the frequency spectrum of the polarized radiations, corresponding to Fig.\ref{fig3}.  \textcolor{black}{In both Fig.\ref{fig3} and Fig.\ref{fig6}, the energy spectra resemble with the synchrotron spectrum. The spectra is concentrated in the region of frequencies extending up to, $\omega/\omega_p \sim B_0 \gamma_0^2$~\cite{Landau:2005fr}. For the plasma density used in our simulations, $n_0 =10^{16}$cm$^{-3}$, this spectrum confirms the existence of the circular polarization in the optical/infra-red frequency regime. One may note that this emitted spectrum is independent of the plasma density. The main role of background plasma is to provide the energy dissipation via the the WI/CFI mechanisms to cause energy difference between the two beam species. Moreover, the inhomogeneous magnetic field generated by the WI/CFI mechanisms, does depend on the beam to plasma density ratio, $n_{b0}/n_0$. This inhomogeneous magnetic field has a de-polarizing effect and it lowers the degree of circular polarization given by Eq.\ref{scaling}~\cite{sinha2019}. Later in Sec.\ref{grbs}, we discuss the relevance of the PIC simulation results on  the circularly polarized radiation in optical frequency range to the observation of circularly polarized radiation in GRBs}. We again repeated the simulations for different beam to plasma density ratios, $n_{b0}/n_0 = [0.1 - 1]$. Figure \ref{fig7} shows the degrees of linear and circular polarizations for a range of beam to plasma density ratios, and the magnetization parameter $\sigma_0$. One can see that the higher Lorentz factor cases (low $\sigma_0$) show negligible amount of the polarised radiation. These plots indicate that the initial $\gamma$ of the plasma jets significantly affect the polarization of the generated radiation.
\textcolor{black}{The reduction in the magnitude of the {circularly} polarized radiation at high $\gamma_0$ is due to two reasons. First, the growth of the {WI/CFI} at high $\gamma_0$,  on account of the relativistic electron mass variation, is rather low. Consequently, the asymmetric energy dissipation, responsible for the generation of the {circularly} polarized radiation,  is not pronounced at high $\gamma_0$. Second, the {Larmor radius} of the gyrating beam electrons in the WI/CFI generated magnetic field is larger than the external magnetic field ($B_0$) scale length. Since the magnetic field generated due to the WI/CFI mechanism is strongly fluctuating, it has a depolarizing effect on the radiation generation. Consequently, the generation of the polarized radiation is low in high $\gamma_0$ regime. In low $\gamma_0$ (or high $\sigma_0$) regime, the growth rate of the CFI/WI is rather high and indeed there is a strong polarized radiation generation as seen in Fig.\ref{fig7}(a). The absence of circularly polarized radiation, in this regime, is linked with the cancellation of both right and left-handed helicities. This cancellation is due to the almost equal contributions of electron and positron species in generating a strong quasi-static magnetic field due to the WI/CFI mechanisms.}

\subsection{Effect of transverse beam size on the polarized radiation generation}

As mentioned earlier, the filamentation of the electron-positron beam in an electron-proton plasma depends on the transverse beam size of the pair beam. If the pair beam has transverse size of plasma skip-depth, $\sim c/\omega_p$, then the beam suffers no filamentation and the the energy difference between electrons and positrons does not arise. Consequently one can not expect generation of circularly polarized radiation since it crucially depends on the energy difference between the electrons and positrons~\cite{sinha2019}. While for pair-beam with larger transverse sizes, one can expect sever filamentation and energy differences between electron and positron species of the beam. Thus, it is important to examine the effect of the transverse beam-size on the polarized radiation generation.

\begin{figure}
\includegraphics[width=0.35\textwidth,height=0.33\textwidth]{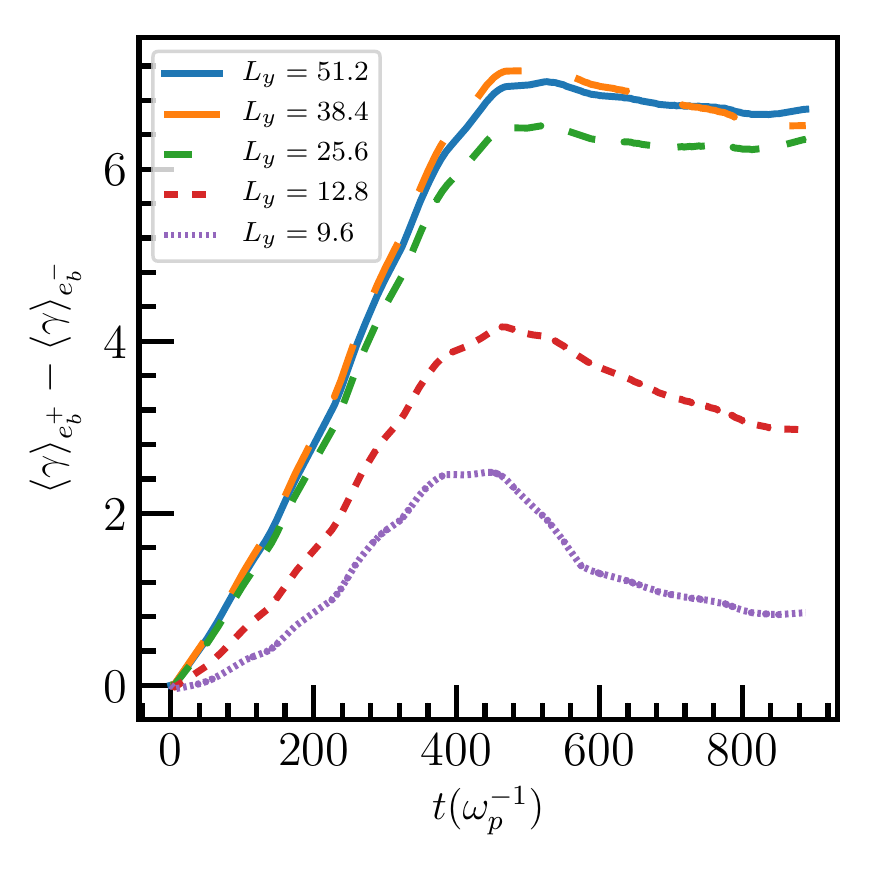}
\caption{Time evolution of the difference in average kinetic energy between the beam positrons, $\langle \gamma\rangle_{e_b^+}$, and the beam electrons, $\langle \gamma \rangle_{e_b^-}$ for varying beam spot size, $L_y$. $L_y$ is given in the units of $c/\omega_p$. The density ratio is $n_{b0}/n_0=1$, and the beam Lorentz factor {$\gamma_0 = 50$}.}
	\label{fig1}
\end{figure}
\noindent
\noindent
\begin{figure}
\centering
\includegraphics[width=0.48\textwidth,height=0.27\textwidth]{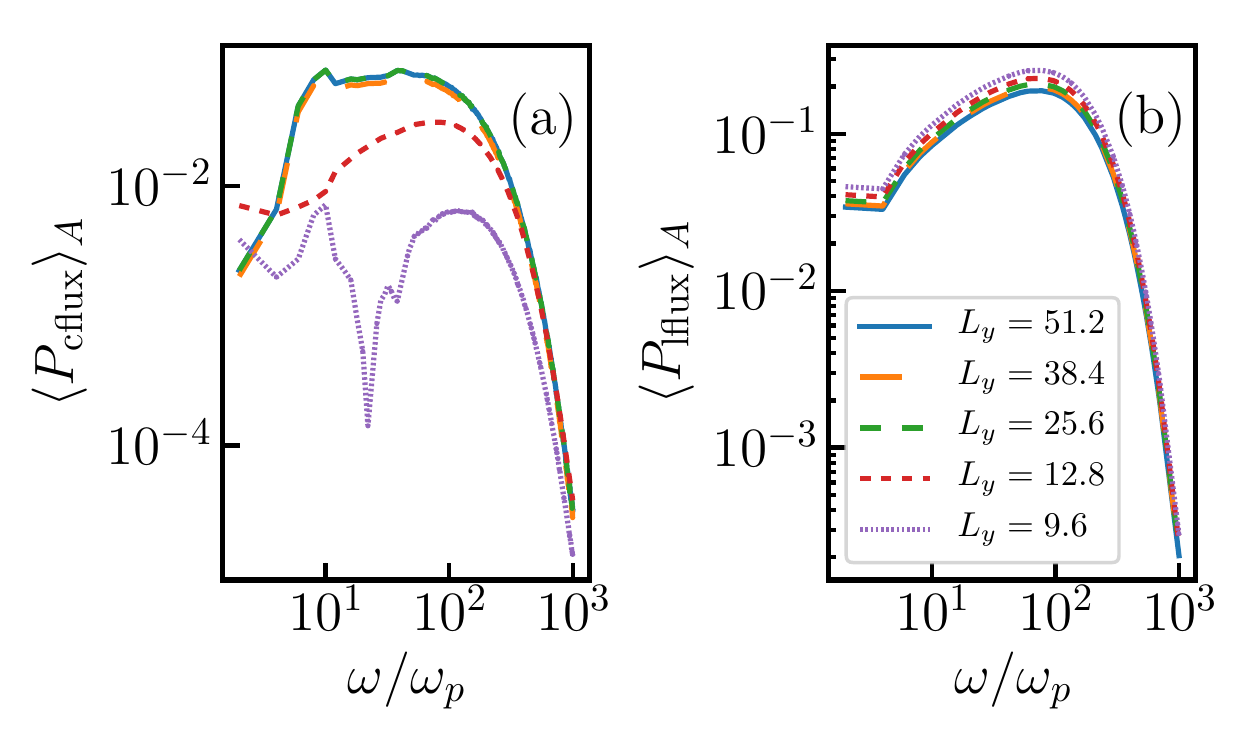}
\caption{(a) The circularly, and (b) linearly polarized radiation fluxes obtained from the beam particles corresponding to Fig.\ref{fig1}. The initial magnetization is {$\sigma_0=0.04$}.}
\label{fig2}
\end{figure}
To investigate this, we keep the longitudinal size of the simulation window constant at \textcolor{black}{$\Delta x=16\,c/\omega_p$ and the transverse size was varied from $\Delta y=\Delta z=[16-64]\,c/\omega_p$.} The resolution along the $x$ direction is $16$ cells per $c/\omega_p$, and $8$ cells per $c/\omega_p$ along the $y$ and $z$ direction are used. Pair beams of longitudinal length \textcolor{black}{$L_x= 6\, c/\omega_p$} and transverse sizes along $y$ and $z$ ranging from \textcolor{black}{$L_y=L_z= [9.6 - 51.2]\, c/\omega_p$} are used. {For all the cases, the beam had a Gaussian density profile in all directions with a FWHM of one-third of the beam length in the corresponding direction.} {This} beam profile resembles a pancake and due to large transverse sizes the oblique mode does not dominate the filamentation dynamics instead the transverse {WI/CFI} mechanisms dominate. Also keeping in mind the available beam-energy in laboratories, we choose {$\gamma_0=50$} and beam to background plasma density ratio  $n_{b0}/n_0=1$. A constant initial magnetic field $B_0 = 2$ is applied which corresponds to magnetization parameter, {$\sigma_0=0.04$}. \textcolor{black}{As shown in Fig. \ref{ene_den_map} before, there is an asymmetry in kinetic energy distribution of both species, with positrons being more energetic than the electrons~\cite{sinha2019}. Figure \ref{fig1} further explores the time development of this particle energy distributions and  shows the difference in the average kinetic energies between the beam positrons ($\langle \gamma\rangle_{e_b^+}$) and the electrons ($\langle \gamma\rangle_{e_b^-}$) for the beams with different spot-size ($L_y$). One can immediately notice that the difference in the energy distributions become larger at larger transverse spot-sizes.} This is expected since larger beam transverse size results in the generation of several filaments, the difference in energy between the positron and electrons increases~\cite{sinha2019}. This difference in beam species energy is spent in generating a quasi-static magnetic field due to the WI/CFI mechanisms. The energy carried into this magnetic field is typically few percent $\le 10\%$ of the initial beam energy, corresponding to the equipartitioning level in this process. With a larger beam spot-size the total energy contained in the beam increases, generating a stronger magnetic field. A stronger magnetic field traps the background plasma particles and consequently it saturates due to nonlinear wave-particle interaction. Thus, no further difference can arise between the {electron} and {positron} energies as seen in Fig.\ref{fig1} for large transverse sizes ($L_y\ge 38.4\, c/\omega_p$). One must also note that the filament merging can also contribute to the saturation with time. Fig.\ref{fig2} shows the frequency spectrum of both circularly and linearly polarized radiations for different beam spot-sizes. The spectrum is calculated by averaging over the detector size of $[30000\times 30000](c/\omega_p)^2$ resolved using $[50\times 50]$ grid points. The frequency range is chosen to be $\omega=[1-10^3]\omega_p$ with the step-size $\Delta\omega=\omega_p$. As expected from Fig.\ref{fig1}, the polarized radiation generation is maximum in the case of a beam with large transverse spot-size and it also saturates for size {$L_y\ge 25.6\, c/\omega_p$}. One {can} again note that the polarized light generation is in optical/infra-red frequency range.

\subsection{Effect of the longitudinal beam size on polarized radiation generation}\label{long_beam}

If the longitudinal size of the beam is large (few tens of $c/\omega_p$), then the coupling of longitudinal electrostatic mode with the transverse {WI/CFI} mechanisms results in the generation of ``oblique mode"~\cite{bret2006,Karmakar:2009aa}. \textcolor{black}{This oblique mode is a manifestation of the transverse and longitudinal modes coupling and is the fastest growing mode in the beam-plasma system~\cite{bret2006}.} The existence of this mode requires the perturbation wavevector to have a component in the direction of the beam propagation, corresponding to the two-stream instability. Also the longitudinal beam length has to be large enough for this mode to grow unlike in Sec.\ref{gamma} where the smaller longitudinal size does not support the excitation of the oblique mode. This oblique mode has the fastest growth rate of any mode of the {WI/CFI} mechanisms in the relativistic regime~\cite{bret2006}. Thus, it is instructive to examine the role of the oblique mode on the polarized radiation generation.
\begin{figure}
\includegraphics[height=0.5\textwidth,width=0.45\textwidth]{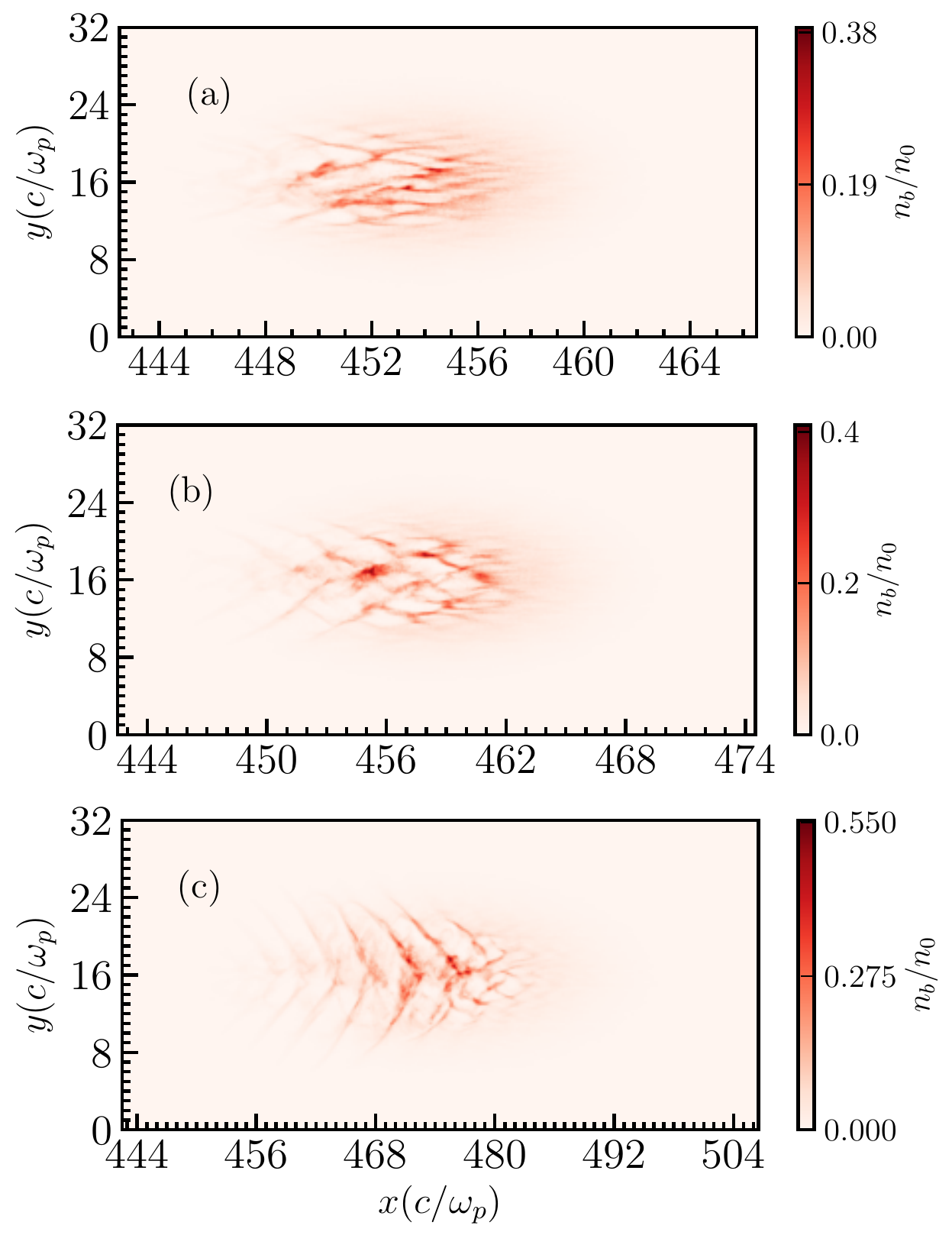}
\caption{Beam electron densities evolution for a ($e_b^+,e_b^-$) beam with density ration $n_{b0}/n_0 = 0.1$, and Lorentz factor {$\gamma_0=50$} in an ambient plasma with magnetization parameter {$\sigma_0=0.4$} at $t = 450\,\omega_p^{-1}$. Panel (a) is for initial longitudinal beam length $L_x = 8\, c/\omega_p$, panel (b) is for  $L_x = 25.6\, c/\omega_p$, and panel (c) is for beam length $L_x=51.2\, c/\omega_p$.}
	\label{fig8}
\end{figure}
\begin{figure}
\includegraphics[height=0.5\textwidth,width=0.45\textwidth]{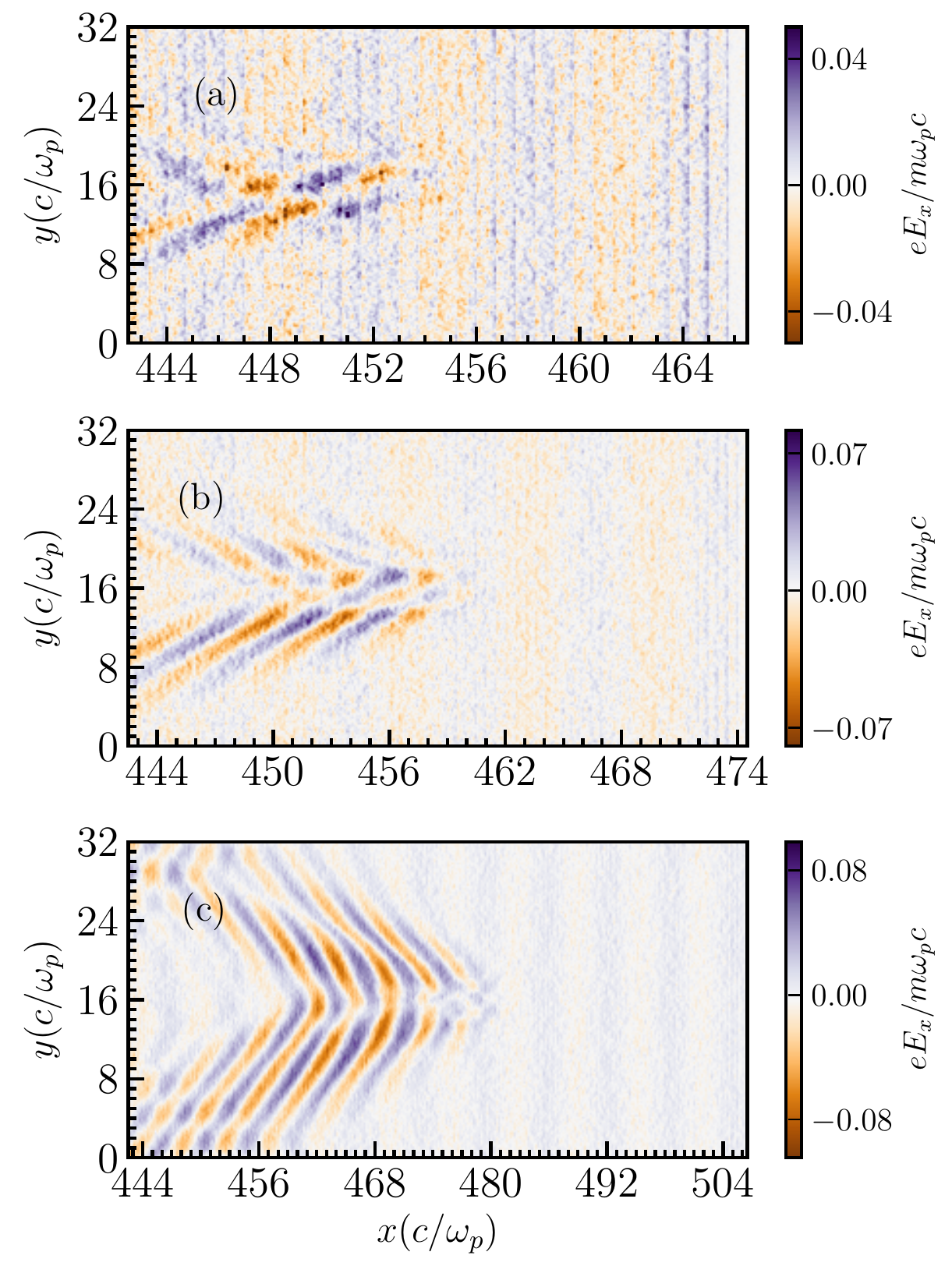}
\caption{Longitudinal electric field $E_x$ for ($e_b^+,e_b^-$) beam with $n_{b0}/n_0 = 0.1$ propagating with {$\gamma_0=50$} in an ambient plasma with {$\sigma_0=0.4$} at $t = 450\,\omega_p^{-1}$. Panel (a) shows  $E_x$ for beam length $L_x=8 c/\omega_p$, (b)  shows  $E_x$ for beam length $L_x=25.6\, c/\omega_p$, and (c) shows $E_x$ for beam length $L_x=51.2\, c/\omega_p$.}
\label{fig9}
\end{figure}
\noindent
To study this, we use a simulation box of transverse size at \textcolor{black}{ $\Delta y=\Delta z= 32\,c/\omega_p$ and the longitudinal size of the simulation window was varied from $\Delta x=[16- 64]\,c/\omega_p$.} The resolution is kept the same as the previous case i.e  $16$ cells per $c/\omega_p$ along the $x$-direction, and $8$ cells per $c/\omega_p$ along the $y$ and $z$ directions, respectively. {The initial beam Lorentz factor is $\gamma_0 = 50$ and the corresponding  magnetization parameter is $\sigma_0 = 0.4$}. In order to carefully analyze the transition from a pure transverse instability to oblique instability, the longitudinal beam length is varied from $L_x=[6-51.2]\, c/\omega_p$ {with the FWHM one-third of the corresponding longitudinal length} while keeping the transverse size constant at $L_y=L_z=25.6\, c/\omega_p$. PIC simulations are performed for a range of beam to plasma density ratios, $n_{b0}/n_0 = [0.1 - 0.5]$. Figure~\ref{fig8} shows the beam electron density at an instant $t=450\,\omega_p^{-1}$ for different beam lengths at a fixed density ratio $n_{b0}/n_0=0.1$. One can immediately notice the transition of filaments structures from purely transverse filamentation caused by the {WI/CFI} mechanisms [panel (a)] to the dominance of the oblique mode [panel(c)], as discussed earlier. For the larger longitudinal beam length, the beam filaments are inclined at an angle confirming the presence of oblique modes. Figure \ref{fig9} shows the associated longitudinal electric fields, $E_x$, with the filaments from Fig.\ref{fig8}. $E_x$ is stronger in the case when the oblique mode is dominant. \textcolor{black}{Since the growth rate of the oblique mode is large, it consequently generates a stronger oscillating  longitudinal electrostatic field, as seen in Fig.\ref{fig9}. The generation of the stronger oscillating electric field can compensate the average energy difference between the {electron} and positron species, leading to an identical energy distribution of both species, which reduces the circularly polarized radiation generation.} Fig.\ref{fig10} shows the time evolution of the average kinetic energy of the beam species for different beam lengths. It is clear that the difference in kinetic energy between the beam species narrows with the onset of oblique instability.
\noindent
\begin{figure}
\centering
\includegraphics[height=0.55\textwidth,width=0.45\textwidth]{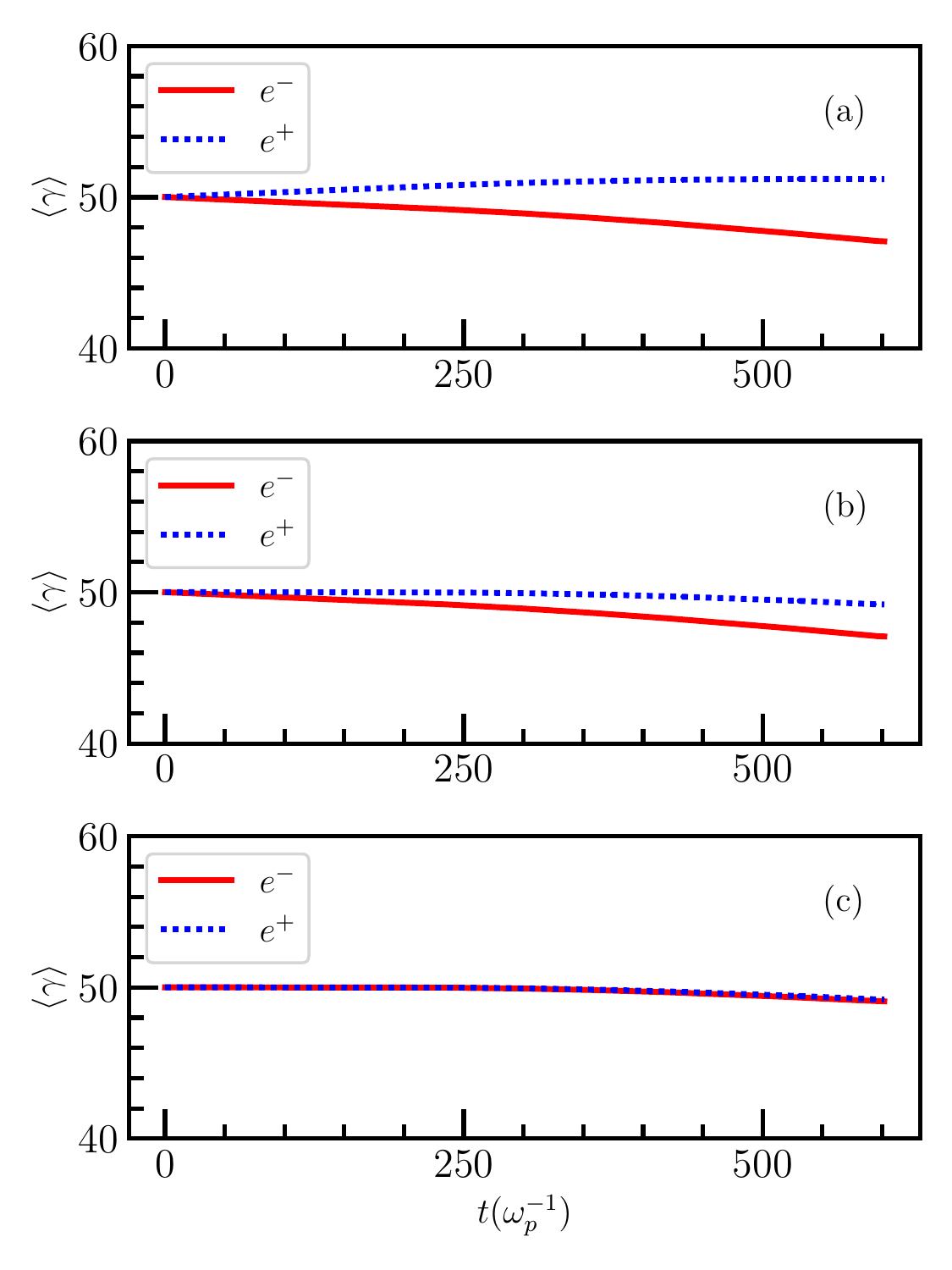}
\caption{Time evolution of the average kinetic energy ($\langle \gamma\rangle$) of the $e_b^+$ (blue, dotted light) and $e_b^-$ (red, solid gery) for $n_{b0}/n_0 = 0.1, \gamma_0 = 50$, and $\sigma_0=0.08$ at $t = 600\,\omega_p^{-1}$. Panel (a), (b) and (c) show the $\langle \gamma\rangle$ for beam length $L_x=8\, c/\omega_p$, $L_x=25.6\, c/\omega_p$, and  $L_x= 51.2\, c/\omega_p$, respectively.}
\label{fig10}
\end{figure}
\noindent
As $\langle P_\mathrm{circ}\rangle \propto (\langle\gamma^4\rangle_+-\langle\gamma^4\rangle_-)/(\langle\gamma^4\rangle_++\langle\gamma^4\rangle_-)$ and the difference in kinetic energy is significantly affected by the onset of oblique modes, it impacts the generation of circularly polarized light flux $\langle P_\mathrm{circ}\rangle$. We extract the trajectories of $1000$ beam particles each in the same way as in Sec.\ref{gamma} and compute the radiation. The detector is placed at a distance of {$10^5\, c/\omega_p$} along the $x$ direction. The detector dimensions are {$10^4\, c/\omega_p$} along $y$ and $z$ directions and resolved using $50\times 50$ grid points and the frequency bandwidth  $[1-10^3] \,\omega_p$ with $\Delta\omega = \omega_p$. Figure \ref{fig11} shows the contour plot  for $\langle P_\mathrm{circ}\rangle$ obtained for beams with lengths, $L_x=[6- 51.2]\, c/\omega_p$ and $n_{b0}/n_0 = [0.1 - 0.5]$. It is clear that the degree of circular polarization decreases with the increase in longitudinal beam length, as discussed before. This observation is of significant importance as it allows to detect the transition from pure filamentation to oblique modes in laboratory astrophysics experiments using the polarized radiation as a diagnostic tool, thus paving a way towards the direct experimental verifications of WI/CFI mechanisms.\\
\noindent
\begin{figure}
\includegraphics[width=0.43\textwidth,height=0.3\textwidth]{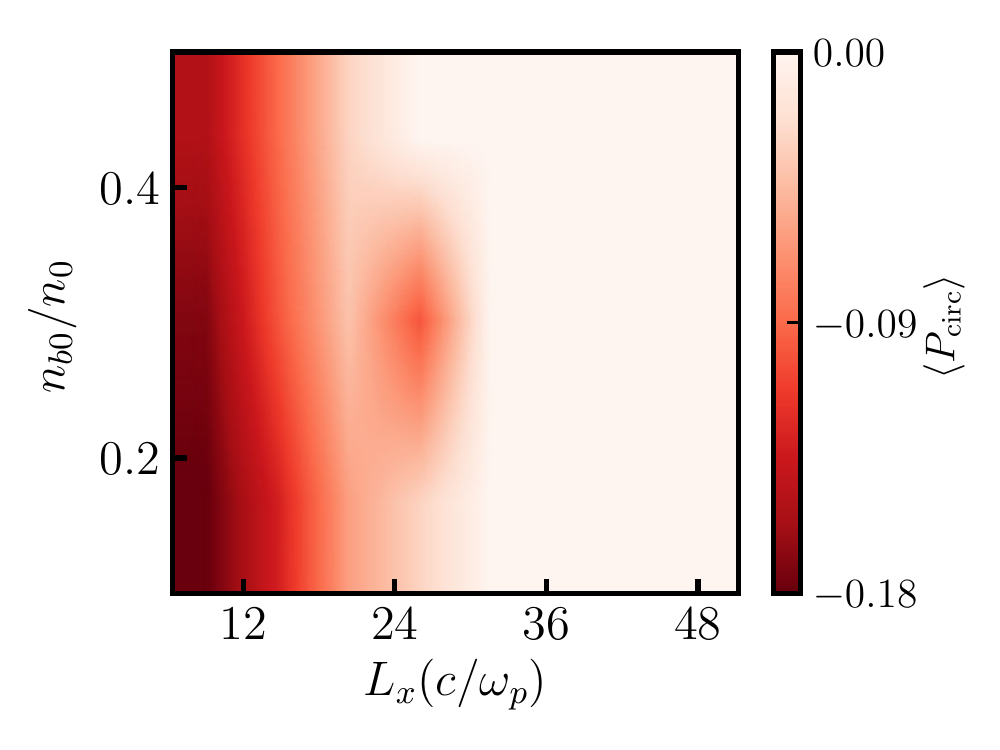}
\caption{Contour plot for the degree of circular polarization, $\langle P_\mathrm{circ}\rangle$, for {$\gamma_0=50$}, $n_{b0}/n_0=[0.1-0.5]$, and $L_x=[6 - 51.2]\,c/\omega_p$.}
\label{fig11}
\end{figure}
\section{Relevance to the observation of polarized radiation in GRBs}\label{grbs}

\textcolor{black}{As GRBs are extremely energetic events, it is reasonable to expect {that} a copious amount of pair plasma is produced along with the presence of baryonic matter in the ejecta into the interstellar medium. This pair-plasma, often termed as the fireball beam, can propagate in a plasma composed of baryonic matter~\cite{piran2005,Anantua:2019aa}. The widespread presence of the magnetic field in the cosmos makes the baryonic plasma to be magnetized.  Specifically the jets in GRBs can be magnetized either due to the magnetic field of the central engine, compressed interstellar magnetic field  at the shock front, or due to the magnetic field generated by other  processes. Thus, modelling the propagation of an electron-positron beam in an external magnetized electron-proton plasma mimics one of the likely scenarios in GRBs. }
To discuss the relevance of our results to the observations of the circular polarizations in GRBs, we invoke the arguments of the scaling and self-similarity between the physical processes in laboratory and astrophysical environments~\cite{Connor:1984aa,Connor:1977aa,Connor:1988aa,Petty:2008aa,ryutov2012}. Connor and Taylor showed that if physical equations of a system are invariant under some scaling transformation, the results derived from these equations must exhibit some scale invariance~\cite{Connor:1977aa}. Thus, if a set of linear transformations leaves the collisionless Vlasov-Maxwell system of equations invariant then we can scale our results to astrophysical scenarios. One can easily verify that the set of following transformations $x \rightarrow a x, t \rightarrow a t, E \rightarrow b E, B \rightarrow b B, \gamma \rightarrow a b\gamma, n \rightarrow (b/a) n$, where $a$ and $b$ are dimensionless constant, $x$ and $t$ are the physical and temporal coordinates, $E$ and $B$ are the electric and magnetic fields, $\gamma$ and $n$ are the Lorentz factor and density, respectively leaves the Vlasov-Maxwell system of equations invariant~\cite{Petty:2008aa,Sarri:2015aa}. It's clear from here that the scaling from laboratory to astrophysical scenarios requires $a \gg 1$ and $b \ll 1$ corresponding to large spatial and temporal scales, and smaller electromagnetic fields and plasma densities, respectively. The spatial and time scales of interest in our case are related to the electron-plasma skin-depth and electron-plasma frequency.  An important quantity which is invariant under the scaling transformation is the magnetization parameter {$\sigma_0=B^2/4\pi\gamma_0 n_{b0} m_e c^2$}. Keeping in mind the fact that we are comparing the polarization of radiation which is a dimensionless quantity, one can employ the scaling principle to extrapolate the significance of our results to observations of circular polarization in GRBs. On applying the scaling argument, the laboratory scale length, $L_\mathrm{l}$ and time $t_\mathrm{l}$ can be related with the corresponding time ($t_\mathrm{A}$) and length ($L_\mathrm{A}$) in the astrophysics as $t_\mathrm{A}=t_\mathrm{l}\sqrt{n_\mathrm{0}/n_\mathrm{A}}$ and $L_\mathrm{A}=L_\mathrm{l}\sqrt{n_\mathrm{0}/n_\mathrm{A}}$, where $n_\mathrm{A}$ is the typical plasma density in the astrophysical scenario.

\begin{table}[ht]
\caption{Characteristic parameters for comparison}
\centering
\begin{tabular}{c c c c}
\hline\hline
Quantities & Laboratory setting &  Astrophysical setting\\ [0.5ex] 
\hline
$n_0$& $10^{16}$ cm$^{-3}$ &1 cm$^{-3}$ \\
$c/\omega_p$&$5.33\times10^{-3}$ cm&$5.33\times10^{5}$ cm \\
$\omega_p^{-1}$&$1.1\times 10^{-12}$ s&$1.1\times 10^{-4}$ s  \\
$B$ & $(3-30)\times 10^{4}$G & $(0.3-3)$mG  \\
$t_\mathrm{sim}$ & $2.66\times 10^{-10}$s & $2.66\times 10^{-2}$s \\ [1ex]
\hline
\end{tabular}
\label{table:scaling}
\end{table}

\textcolor{black}{Table~\ref{table:scaling} shows the scaling of the characteristic parameters from PIC simulations to the astrophysical scenarios in a comoving frame where the source (plasma jet) is at rest~\cite{piran2005,Sagiv:2004aa}. The simulation time in our case is within the  time range $(0.01-100)$s reported for GRBs~\cite{piran2005}. We use a range of the scale invariant parameter $\sigma_0 \in{[0.002,0.2]}$ in PIC simulations. For the internal shocks scenario in GRBs, this range of $\sigma_0$ can be similar to the range reported elsewhere~\cite{Waxman:1997aa,sironi2009A,harrison2013}, though for the forward shock regions the magnetization levels are smaller than $\sigma_0$~\cite{lemoine2013}. The longitudinal beam sizes  considered in Sec.\ref{long_beam} for the  astrophysical plasma density, $n_0 \approx 1\,\mathrm{cm}^{-3}$, are in the range $[3.2\times 10^6 -1.4\times 10^7]$ cm, which is smaller than the width of the relativistic shell in GRBs~\cite{piran2005}. Thus, the WI/CFI mechanisms have ample spatial extent and time to grow and affect the polarized radiation generation. 
Hence, our choices of the beam plasma parameters, on scaling, covers a broad range of parameters relevant for GRBs and therefore our results, upon scaling, can help in understanding the generation of polarized radiation in GRBs. }

\begin{figure}
\includegraphics[width=0.42\textwidth,height=0.3\textwidth]{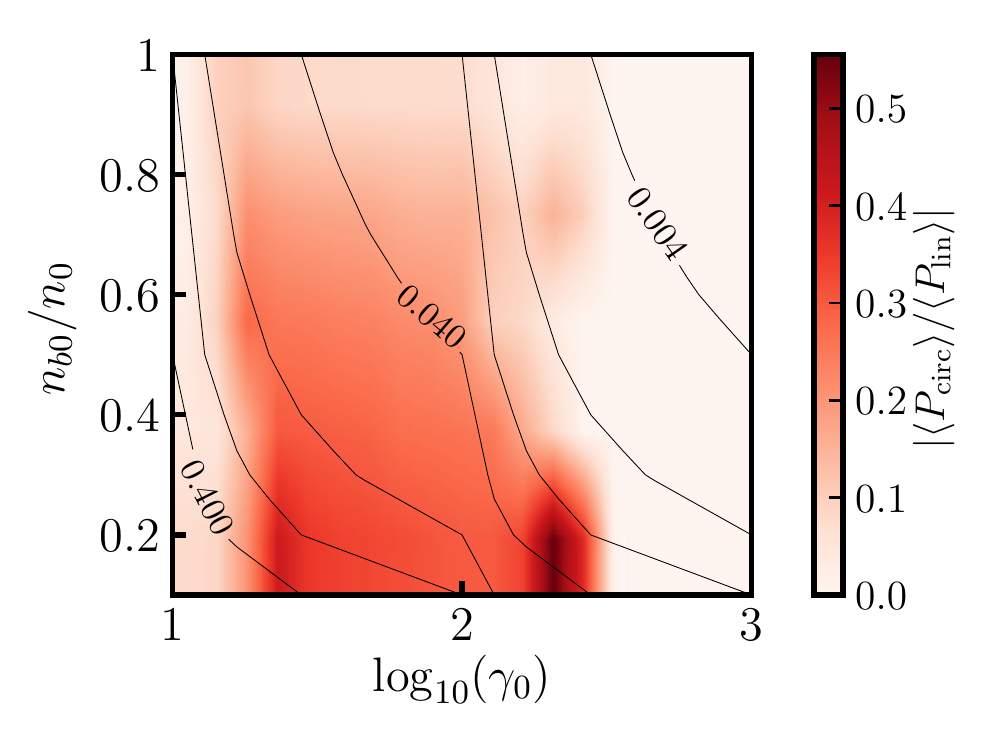}
\caption{Parameter map for {$|\langle P_\mathrm{circ}\rangle/\langle P_\mathrm{lin}\rangle |$} corresponding to {Fig.\ref{fig7}. The overlaid contour lines depict the magnetization parameter $\sigma_0 \in [0.002,0.2]$}.}
\label{fig12}
\end{figure}

The observed spectrum of the radiation is synchrotron like in GRBs as in the PIC simulation results~\cite{sinha2019}. According to the theory, this synchrotron radiation is supposed to be linearly polarized in the observation direction. Hence, the detection of circular polarization in GRB 121024A is particularly interesting. A finite degree of circular polarization can be attributed to either the anisotropy in the pitch angle distribution, or the off-axis radiation observations. Recently, it has been shown that neither of these factors can account for the observations of the circular polarization for a realistic scenario~\cite{nava2015}. 
\textcolor{black}{As mentioned before in Sec.\ref{gamma}, PIC simulations show the generation of circular polarization in the optical frequency range. On scaling the parameters to the astrophysical scenario from Table~\ref{table:scaling}, one must also take into account the bulk Lorentz factor of the jets in GRBs~\cite{Ghirlanda:2018aa,Nava2017}. The bulk  motion can not only blueshift the photon emission and but it can also compress the parallel magnetic field along the flow (as in the PIC simulations)~\cite{piran2005}. Thus, the frequency of emission, considered in Sec.\ref{gamma}, is modified as $\omega/\omega_p^A \sim B_0 \Gamma \gamma_0^2$, where $\omega_p^A$ is the local astrophysical plasma density~\cite{piran2005}.  Based on these considerations, the generation of the circular polarization in PIC simulations can be compared with the observation of circular polarization in GRB 121024A~\cite{wiersema2014}.}
Based on our simulation results, we estimate the ratio of the degree of circular to linear polarization for the cases presented in Sec.~\ref{gamma}. We found that a combination of broad range of initial magnetization and density ratios $n_{b0}/n_0$ can yield the polarization levels reported from GRB 121024A~\cite{wiersema2014}. Figure \ref{fig12} shows the ratio of the degree of circular to linear polarizations for the cases described in {Fig.\ref{fig7}}. We observe that the polarization levels depend on the initial magnetizations and the beam to plasma density ratios.  In addition, the plasma is composed of electrons, positrons, and heavy ions. The heavy ions are essential to facilitate an asymmetry in the kinetic energy of the beam species. It has long been argued that the symmetry break in the configuration of the magnetic field is essential for polarized radiation. Our studies show that though the symmetry breaking is necessary which is due to the initial magnetization, this is not sufficient to obtain a polarized radiation. In addition to the initial magnetization there are three conditions that need to be satisfied. First, the plasma must be also composed of baryonic matter \emph{e.g.} electrons, positrons, and ions. Second the ratio of Larmor radius to characteristic size of the CFI/WI induced filaments,  $\eta < 1$. And lastly, the plasma jet should have a geometry such that the effect of the longitudinal mode on the filamentary dynamics is negligible. \textcolor{black}{Thus, our results can shed light on the lack of the circular polarization observations in other GRBs and help in understanding the fireball beam dissipation dynamics in GRBs.}

\textcolor{black}{We also wish to point out that we have not considered the influence of the electron-position beam temperatures on the polarized radiation generation in this paper. The transverse momentum spread in the beam distribution function can suppress the CFI/WI type transverse instabilities~\cite{bret2006,Karmakar:2009aa}. However, the oblique mode still persists in this scenario for a broad range of parameters~\cite{Karmakar:2009aa,Bret2010}. Since the oblique mode seems to reduce the generation of polarized radiation, the lack of circularly polarized radiation observation in other GRBs can be attributed to the combined effects of the longitudinal size and the momentum spread of the fireball beam. However, the beam and plasma temperatures can also affect the hierarchy of beam-plasma modes~\cite{Bret2010}. Thus, on tuning of the beam-plasma temperatures, the observation of circular polarization even for large  longitudinal beam sizes can still be possible in a GRB scenario. }

\section{Conclusions}\label{conclusions}

In this paper, we have estimated the levels of linear and circular polarizations from the ($e_b^+,e_b^-$) beam species during its interaction with an ambient magnetized electron-proton plasma. The radiated fields and the corresponding degrees of polarization are calculated numerically by post-processing a sample of trajectories of the beam species extracted directly from PIC simulations. We have studied the polarization properties by varying both the $\gamma_0$ (consequently the magnetization parameter $\sigma_0$) and the transverse and longitudinal lengths of the pair-beam. When the transverse beam size is few tens of $c/\omega_p$ and it is larger than the longitudinal size (few $c/\omega_p$), the filamentation of the pair-beam is dominated by the  WI/CFI mechanism. In this case, we obtain a significant amount of both linear and circular polarizations for a beam with varying Lorentz factors. With higher $\gamma_0$, the Larmor radius of electrons in a magnetic field increases. Since the WI/CFI generated magnetic field is inhomogeneous and fluctuating, it has a strong de-polarizing effect of the polarized radiation generation. Hence, the levels of linear and circular polarizations decrease with $\gamma_0$. In the case of longitudinal beam length being larger than transverse length, the oblique mode associated with WI/CFI mechanisms dominates. The oblique mode gives rise to strong electrostatic field, providing periodic acceleration and deceleration zones for both beam species. Thus, the difference in kinetic energy between the beam species reduces with the longitudinal beam lengths, reducing the generation of radiation with circular polarization.

Based on these observations, one can imagine that the origin of polarized radiation in GRBs, in particular the circular polarization,  is a result of the combined effects of large scale ordered magnetic field, plasma composition, initial $\gamma_0$ (hence $\sigma_0$), and the transverse extent of the fireball beam in plasma jets. A combination of these parameters can yield the value of $\langle P_\mathrm{circ}\rangle /\langle P_\mathrm{lin}\rangle=0.15$, which is reported in the GRB 121024A~\cite{wiersema2014}. \textcolor{black}{One may note that the observation of the circularly polarized radiation in the optical frequency range comes from the afterglow emission in GRBs. It is presumed to occur in the forward shock region which has a lower magnetization level than the range  $\sigma_0 \in{[0.002,0.2]}$ considered in PIC simulations.} 
Our results can also be significant for ongoing and planned experiments to investigate the dynamics of the WI/CFI mechanisms of a pair-beam. Especially the degree of circular polarization generation in optical/infra-red frequency spectrum can be used to establish the transition  from transverse to oblique modes of the beam-filamentation dynamics, concurrently with the proton-radiography technique.

\appendix
\section{Benchmarks for CASPER code}

\begin{figure}
\includegraphics[width=0.35\textwidth,height=0.45\textwidth]{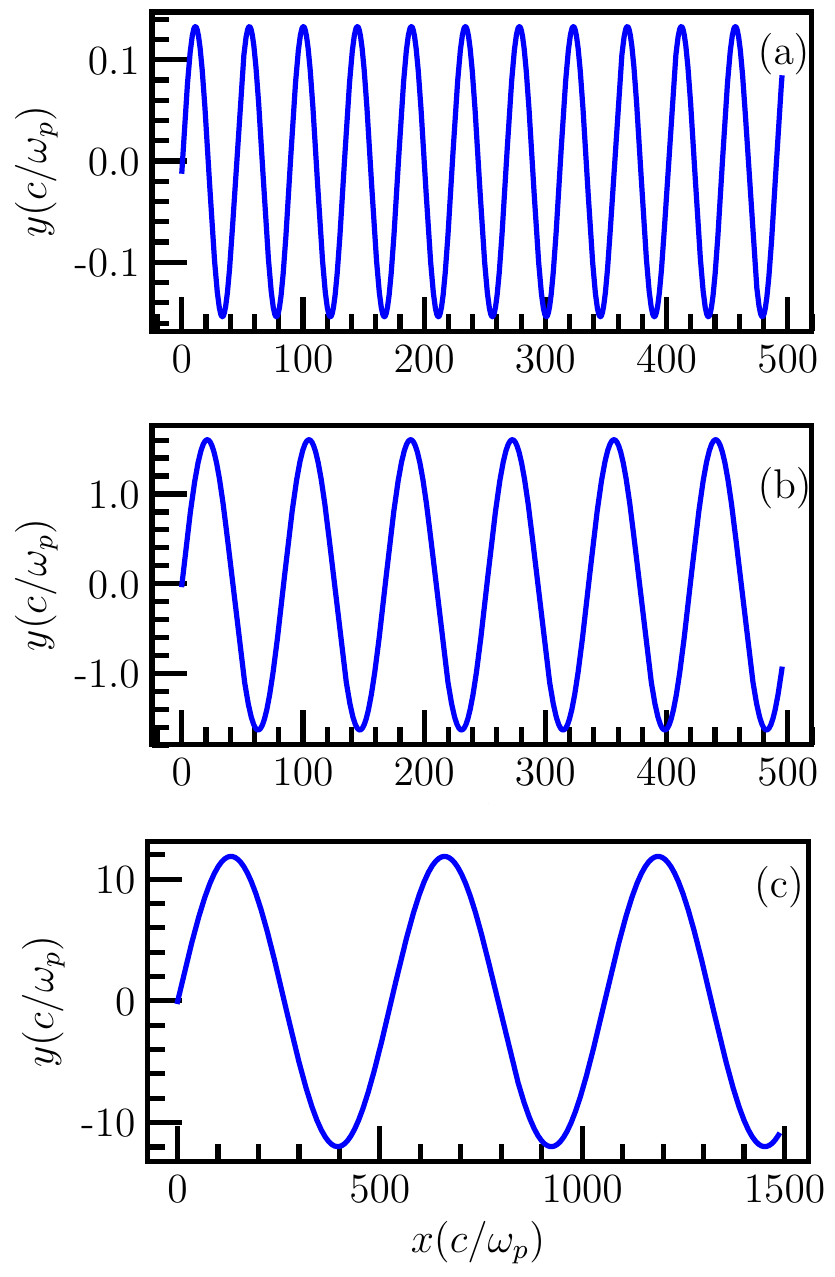}
\caption{Trajectory of a monoenergetic electron beam moving in an ion cavity with wiggler parameter (a) $K = 0.145$, (b) $K = 1.066$, and (c) $K = 2.845$.}
\label{ionchanneltracks}
\end{figure}

We performed several tests to benchmark the radiation code CASPER. These tests were performed for the motion of a monoenergetic electron beam in an ion channel and for the cyclotron motion of a similar beam in a constant magnetic field perpendicular to its velocity. For the motion of the electron in an ion channel, the initial Lorentz factor $\gamma$ was varied to obtain trajectories of different wavelengths, $\lambda_u$ and amplitude $r_u$. Figure~\ref{ionchanneltracks} shows three different trajectories for the electron with $\gamma = [7.2, 8.75, 20.0]$. The corresponding amplitudes and wavelengths are $r_u = [0.145, 1.60, 12.0]c/\omega_p$, and $\lambda_u=[45, 82.5, 530]c/\omega_p$, respectively, where $\omega_p$ corresponds to the plasma frequency associated with the density of the electron beam. Grid resolutions of $32$ cells per $c/\omega_p$ in each direction was used. The trajectories of the electron beam are similar to motion of beam electrons in an undulator or a wiggler. The strength parameter, $K$, associated with these trajectories are $K = [0.145, 1.066, 2.845]$ respectively.  The radiation is emitted at frequencies given by,

\begin{figure}
\includegraphics[width=0.32\textwidth,height=0.5\textwidth]{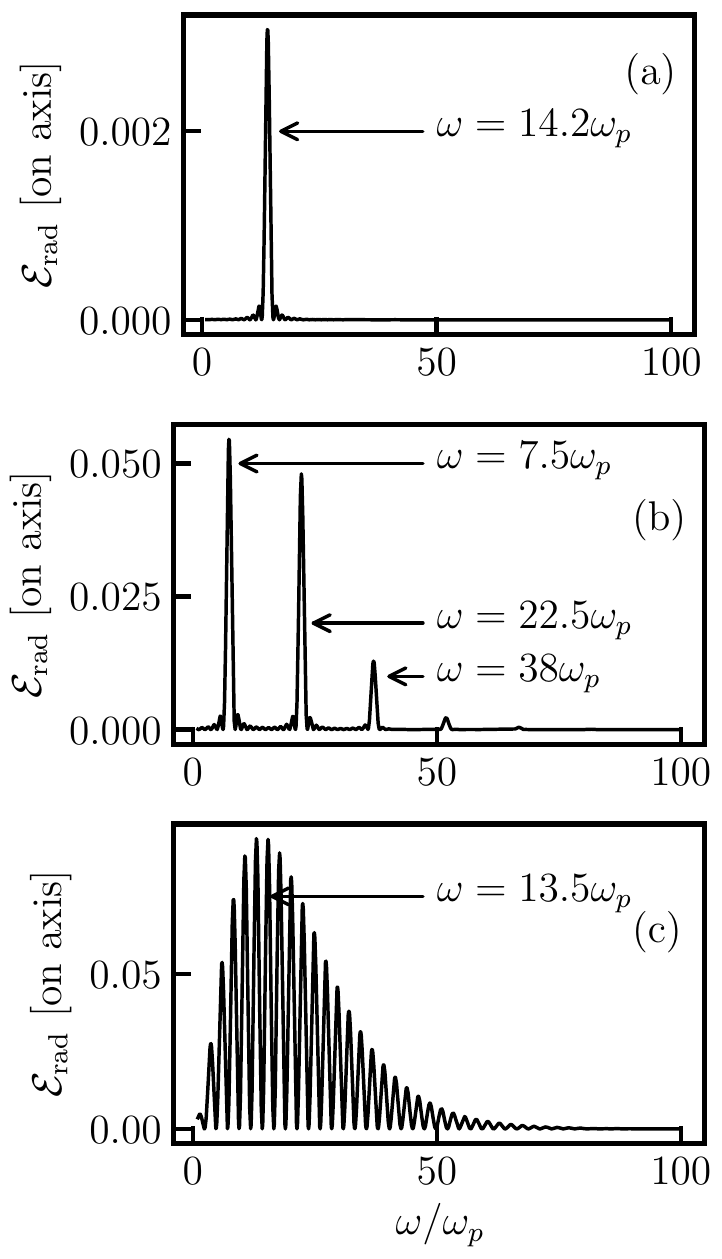}
	\caption{On-axis radiated energy spectra from a monoenergetic electron beam moving in an ion cavity with wiggler parameter (a) $K = 0.145$, (b) $K = 1.066$, and (c) $K = 2.845$.}
	\label{ionchannelspectra}
\end{figure}
\begin{equation}
	\omega_m(\theta) = m\frac{2\gamma^2}{1+K^2/2+\gamma^2\theta^2}\frac{2\pi c}{\lambda_u},\label{undulator}
\end{equation}
where $\theta$ is the angle of radiation emission, and $m$ takes on integer values. The on-axis radiation spectra ($\theta = 0$) for $K=[0.145, 1.066]$ are shown in Figs. \ref{ionchannelspectra} (a) and (b), respectively. For $K = 0.145$, the radiated energy peaks at a single frequency $\omega = 14.5\omega_p$, and for $K = 1.066$ odd harmonics of fundamental frequency $\omega=7.5\omega_p$ are obtained. These values are in perfect agreement with those obtained from Eq.\eqref{undulator}. For $K>1$, the amplitude of the electron motion is, as seen in Fig.\ref{ionchannelspectra}(c), is large and the trajectory exhibits the characteristics of a wiggler. The radiation emitted in this case has a broad spectrum. The frequency at which the peak spectral intensity is obtained is given as
\begin{equation}
	\omega_c = \gamma^2K\frac{2\pi c}{\lambda_u}.
	\label{wiggler}
\end{equation}
For the case in Fig.\ref{ionchanneltracks}(c), Eq.\eqref{wiggler} yields $\omega_c = 13.5\,\omega_p$. This is also confirmed in Fig.\ref{ionchannelspectra}(c) obtained from CASPER.
\begin{figure}
	\includegraphics[width=0.55\linewidth]{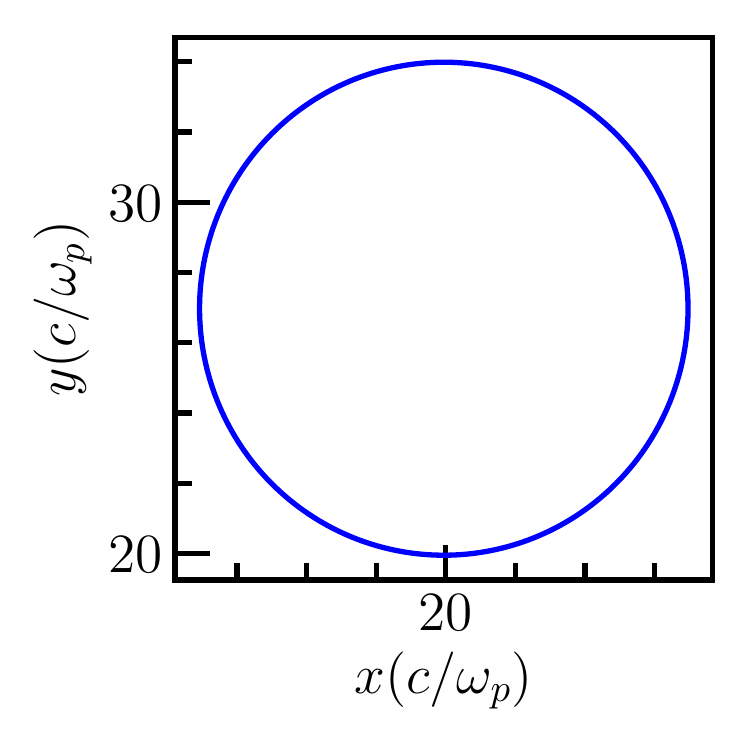}
	\caption{Trajectory of a monoenergetic electron beam with $\gamma = 7$ in a uniform static magnetic field $B_z=1$ along the \textcolor{black}{z} $x$ direction.}
	\label{cyclotrontrack}
\end{figure}
\begin{figure}
	\includegraphics[width=0.38\textwidth, height=0.25\textwidth]{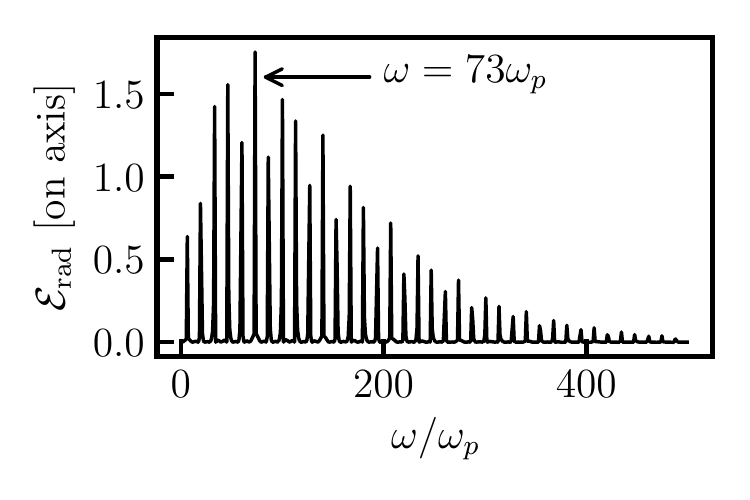}
	\caption{On-axis radiation spectra from the electron beam shown in Fig.\ref{cyclotrontrack}.}
	\label{cyclotronspectra}
\end{figure}

Next we analyze the radiation from the cyclotron motion of the monoenergetic electron beam in an external magnetic field. The trajectory of the beam is shown in Fig.\ref{cyclotrontrack}. The beam has a Lorentz factor $\gamma=7$ and the magnetic field of magnitude $B_z = 1$  (normalized to $m_e\omega_pc/e$) is applied in the $z$ direction. We use the same grid resulotion of $32$ cells per $c/\omega_p$ in each direction as before. The radius of curvature is $\rho = 7 c/\omega_p$, and the critical frequency for the radiation from such a motion is given as,
\begin{equation}
	\omega_c = \frac{3}{2}\frac{c}{\rho}\gamma^3.
	\label{cyclotron}
\end{equation}
\begin{figure}
	\includegraphics[width=0.48\textwidth, height=0.3\textwidth]{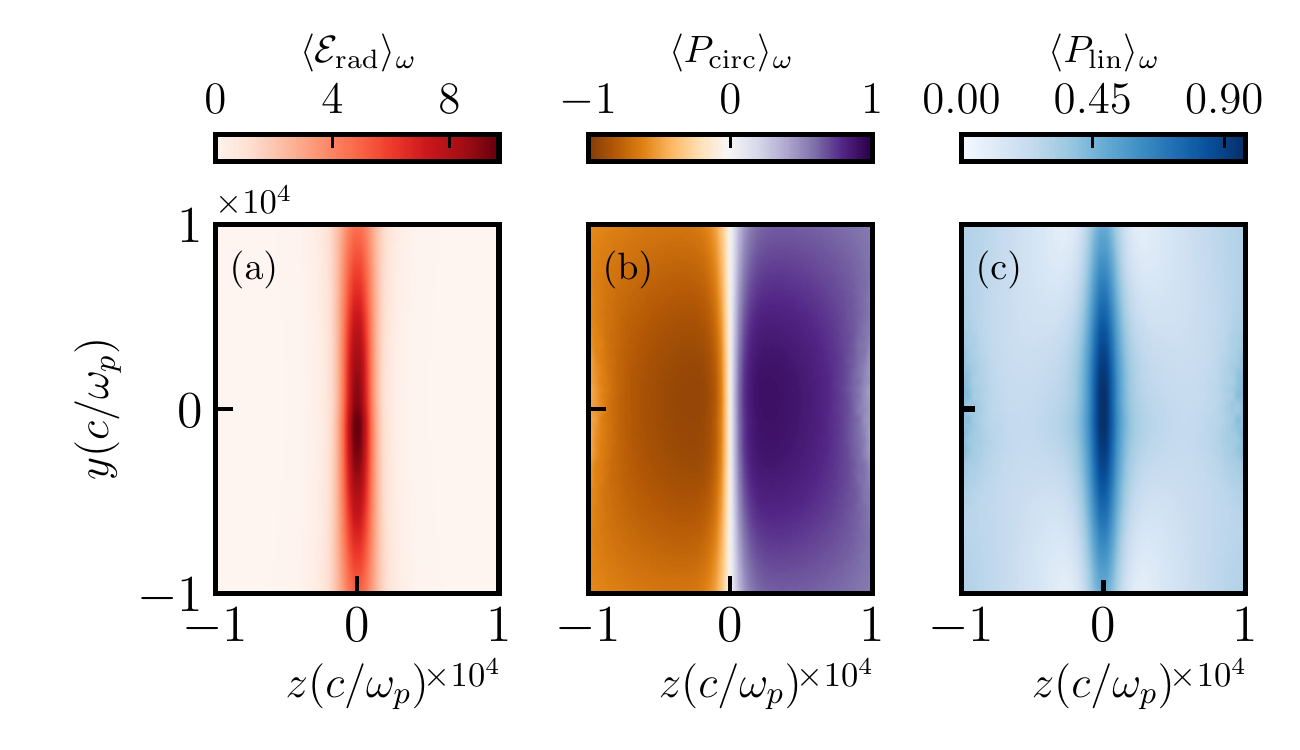}
	\caption{(a) Spatial distribution of the frequency averaged radiated energy, (b) the degree of circular polarization, and (c) the degree of linear polarization from an electron beam shown in Fig.\ref{cyclotrontrack}.}
	\label{freavgspec}
\end{figure}
\noindent
On using Eq.\eqref{cyclotron}, one gets $\omega_c = 73\,\omega_p$. Figure \ref{cyclotronspectra} shows the on-axis radiation spectrum due the cyclotron motion, and the critical frequency, $\omega_c$ is in perfect agreement with Eq.\eqref{cyclotron}. The  radiation from a relativistic accelerating charged particle is emitted in a cone of half angle $\theta = 1/\gamma$. Thus, to capture the total radiated energy, one must measure the radiation at different locations on a plane. This is done by employing a two-dimensional virtual detector which is divided into grid points. To capture the entire radiation emitted due to the cyclotron motion, we use a detector in the $yz$ plane kept at a distance of $10^4 (c/\omega_p)$. The detector has size of $20000\times 20000 (c/\omega_p)^2$ and is divided into $50\times 50$ grid points. The radiated energy and the Stokes parameters are calculated at each of the grid points for a frequency range of $[1-100]\,\omega_p$ resolved using $1000$ frequency bins. The frequency averaged radiated energy, degree of circular polarization, and the degree of linear polarization are shown in Fig.\ref{freavgspec}. As the relativistic particle motion is confined to the $xy$ plane, the radiation is primarily emitted along the $y$ axis. A strong linear polarization observed in the region of intersection of the plane of the particle trajectory and the detector. As the magnetic field points in the positive $z$ direction, the electron rotates in counter-clockwise direction. Hence, on viewing from the positive half of the $z$ axis of the detector, a right circular polarization is observed. Conversely, left circular polarization is observed in the region of the negative $z$ axis of the detector. These observations are in perfect agreement with the theory. 



%

\end{document}